\documentclass[10pt, twocolumn, journal]{IEEEtran}
\usepackage{epsfig}
\usepackage[ruled, lined, longend, linesnumbered]{algorithm2e}
\usepackage{verbatim}
\usepackage{ulem}
\usepackage{amsmath}
\usepackage{textcomp}
\usepackage{mathtools}
\usepackage{graphicx}
\usepackage{longtable}
\usepackage{tablefootnote}

\ifx\pdfoutput\undefined
\usepackage[dvips]{color}
\else
\usepackage{color}
\fi

\renewcommand{\baselinestretch}{1.0}
\usepackage[bookmarks=false]{hyperref}
\hypersetup{%
  unicode   =false,
  pdftoolbar=false,
  pdfmenubar=true
}

\begin{document}
\pagestyle{empty}

\setlength\parindent{0pt}
\setlength\textheight{24.9cm}

\makeatletter
\def\ps@headings{%
\def\@oddhead{\mbox{}\scriptsize\rightmark \hfil \thepage}%
\def\@evenhead{\scriptsize\thepage \hfil \leftmark\mbox{}}%
\def\@oddfoot{}%
\def\@evenfoot{}}
\makeatother

\title{Resource-Efficient Seamless Transitions For High-Performance Multi-hop UAV Multicasting}
\thispagestyle{empty}
\author{Wanqing Tu \\
Department of Computer Science, Durham University, United Kingdom. \\
Email: wanqing.tu@durham.ac.uk}
\maketitle
\pagestyle{empty}\thispagestyle{empty}
\begin{abstract}
Many UAV-related applications require group communications between UAVs to reliably and efficiently deliver rich media content as well as to extend line-of-sight coverage between sky and ground. This paper studies fast yet resource-efficient UAV transitions while maintaining high multicasting performance. We develop a set of analytic and algorithmic results to form the efficient transition formation (ETF) algorithm that deals with different UAV transition scenarios in a multicasting environment. The ETF algorithm first evaluates the seamlessness of a straight-line trajectory (SLT), by processing low-complexity computations (e.g., Euclidean distances) or a chain of fast checks with controlled traffic overheads. For an interrupted SLT, ETF  establishes a new trajectory consisting of a minimum number of seamless straight lines that join at specially selected locations in terms of controlling mobile UAVs' seamless travel distances. Our simulation studies quantify the multicasting performance gains that ETF allows, outperforming compared studies when seamlessly transiting UAV group members.
\end{abstract}
\section{Introduction}
Unmanned Aerial Vehicles (UAVs) are popularly used for acquiring videos or images beyond the capability of traditional ground devices. Aerial multicasts that consist of paths with multiple UAV hops are essential for aerial multimedia group communications in which videos/images are relayed for large-scale ground groups that have members otherwise cut off by obstacles. This is because multicasting allows single UAVs to send/forward the same data only once when distributing the data to a group of receivers, while unicasting processes the same data multiple times when generating/transmitting copies for different receivers and broadcasting floods data among UAVs regardless of whether they are receivers or not. UAV multicasts often need to accommodate UAV movement so as to enable activities such as adapting to the movements of ground devices, capturing different videos/images, etc. Existing studies on UAV transitions in multicast (e.g., [7-8]) mostly investigate UAV positioning or trajectory planning for air-to-earth developments in which UAVs distribute data to a group of ground devices via single hops. UAV transitions in a multi-hop aerial multicast system are rarely studied in the literature. This paper will be among the first to explore the support of such UAV transitions.

Multi-hop multicasts with mobile nodes have been studied for terrestrial multimedia group applications (e.g.,[1-5]). Many of these establish a tree or mesh architecture. When nodes move, tree or mesh architectures need to repair interrupted connections, which requires multicasting nodes to exchange conversations in addition to those issued by the multicasting algorithms. Extra conversations not only consume the energy of wireless nodes but also incur traffic overhead. Geographic multicasts [3-5] reduce such interruption repairs by establishing a tree or mesh topology between zones instead of multicasting nodes. The disconnecting of zones from a tree or mesh is much less frequent than the movement of wireless nodes, and hence fewer connection repairs are needed. However, the formation of zones and the allocation of nodes to different zones involve complex procedures/calculations. This complexity consumes energy and incurs delays. Also, the maintenance of zones caused by nodes' joining or leaving results in extra messages on the multicasting system. Therefore, existing multi-hop multicast algorithms have different drawbacks in supporting node mobility while maintaining multicasting performance. When UAV multicasts adopt these existing solutions, as battery-powered UAVs consume energy more quickly than terrestrial mobile nodes in order to remain airborne, the drawbacks negatively affect both transiting and non-transiting UAVs in the system more severely because the lack of energy hinders interruption repairs, zone maintenance, and timely data distribution. We are hence motivated to design a new approach not only for transiting mobile UAVs with high performance but also avoiding the degradation of performance for other multicasting UAVs.
\begin{figure}[h]
\begin{center}
\begin{tabular}{c}
\includegraphics[trim=530 380 60 130,clip,height=2.2in]{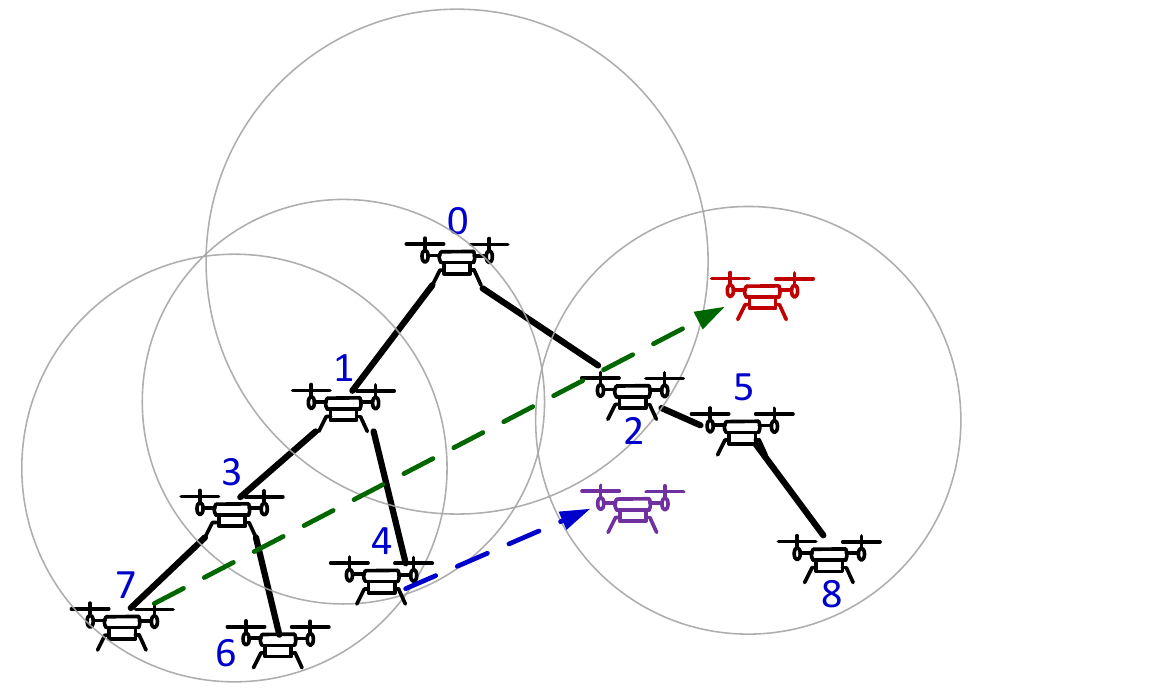}
\end{tabular}
\end{center}
\caption{An example to illustrate seamless transitions in our system.}\label{introduction}
\end{figure}

In response to this motivation, a major focus of our design is on avoiding multicasting architecture repair and controlling additional message exchanges between UAVs when they move. This is because these attributes help to simplify UAV processes and reduce traffic overheads between UAVs, benefitting the maintenance of multicasting performance. Imagine that a UAV knows a set of forwarders which can form a chain of overlapping coverage connecting its origin to destination. This UAV is able to conduct seamless transitions by spontaneously passing between these forwarders without requiring any adjustment of multicasting architecture. By seamless transitions, we mean that UAVs do not lose connections to the multicasting system during their movements from origins to destinations. We use an example in Fig.~\ref{introduction} to illustrate this idea. The coverage of UAVs 3, 0, and 5 overlaps one by one, forming a chain of overlapping coverage. UAV 7 may transit to its destination (highlighted by the red UAV) seamlessly by moving through this chain of overlapping coverage, without adjusting the multicasting architecture.

When developing such a seamless trajectory passing through a chain of overlapping coverage formed by multicasting UAVs, given that we anticipate an energy-efficient solution, the trajectory should be obtained by a low-complexity procedure without requiring UAVs to send/forward extra messages. More importantly, the adopted trajectories should also have travel distances as short as possible. This is because UAVs consume energy quickly during flying. Obviously, a straight-line trajectory (SLT) (e.g., the green dotted arrow line in Fig.~\ref{introduction}) enables a UAV to travel the shortest distance between two locations. The aerial communication environment actually provides a unique advantage for UAVs to transit via an SLT\footnote{In practice, UAVs may experience obstacles. UAV obstacle avoidance is a well-studied topic (e.g., [24]) which may be employed by our UAVs to detour around obstacles and continue transiting via planned trajectories. When we talk about straight lines, we exclude such obstacle-avoidance detours.} because of limited obstacles. However, SLTs are not always seamlessly covered by multicasting forwarders. For example, the blue dotted arrow line in Fig.~\ref{introduction} for UAV 4 to move to the location highlighted by the purple UAV. This leads us to explore the following two research questions.
\begin{itemize}
\item How to determine whether an SLT is seamlessly covered by a set of forwarders in the multicasting system or not?
\item If an SLT is not seamless, how to find the trajectory that offers a short and seamless travel distance passing through the combined coverage of multiple multicasting nodes?
\end{itemize}

To answer the above questions, based on the travel distances of UAVs, we classify transitions into a class with starting and destination locations covered by two forwarders having overlapping coverage, and a class with starting and destination locations covered by two forwarders without overlapping coverage\footnote{Whether two coverage is overlapping or not is defined in Section III.A}. Without loss of generality, we refer to transitions between locations covered by two forwarders having overlapping coverage as short-distance transitions, and to transitions between locations covered by two forwarders without overlapping coverage as long-distance transitions. By studying these two types of transitions, we present the following major new results. These results are then combined to form the efficient trajectory formation (ETF) algorithm that enables UAVs in aerial multicasts to adjust their locations in a fast and performance-guaranteed manner.
\begin{itemize}
\item The seamlessness condition for SLTs in short-distance transitions. This condition requires a UAV to calculate only Euclidean distances when determining whether an SLT is seamless or not, incurring low computational complexity. Also, all information needed by the calculation can be collected using messages already required by multicasting algorithms, avoiding additional message exchanges in the system.
\item Efficient seamless trajectory formation for short-distance transitions. When an SLT is interrupted, it establishes a seamless trajectory with controlled distance passing through the combined coverage of the two forwarders covering the starting and destination locations. Such short seamless trajectories can be achieved by simple computational tasks such as solving quadratic equations, without sending new messages across the multicasting system. This formation scheme improves our study in [17] so that trajectories established can always be seamless.
\item The efficient seamlessness checking algorithm for long-distance transitions. Our previous work [17] does not consider that a long-distance SLT could be seamless. This paper proposes the efficient seamlessness checking algorithm to guarantee that a seamless SLT will not be missed if such an SLT exists. The algorithm refers to a minimum number of multicasting forwarders when deciding whether the combined coverage of multiple forwarders may cover all parts of a long-distance SLT or not. This benefits a fast seamlessness check. Also, the implementation of this algorithm does not require additional message exchanges.
\item The efficient seamless trajectory formation algorithm for long-distance transitions. When an SLT is interrupted, it establishes a new trajectory that enables a UAV to seamlessly transit with a controlled travel distance. Such a new trajectory is formed by a transition UAV that utilises the combined coverage of a minimal number of forwarders without giving rise to additional processes or messages on the multicasting system. As compared to our study in [17] that always checks all multicasting forwarders to seek a path, UAVs in this paper may only check a subset of suitable multicasting forwarders, benefitting a fast start of transitions.
\end{itemize}

Finally, we use NS2 simulations to evaluate our ETF. By observing the average delays and throughput for transiting UAVs and multicasts, and as compared to the studies developed by other researchers, ETF can admit up to 66\% more multicasting traffic with acceptable performance at both transitional and non-transitional UAVs. Our simulations also report that significantly lower control traffic and energy consumption are generated by ETF than by those of related studies.

\section{Related Studies}
Many wireless multi-hop multicasts are designed for group applications with static members. They mainly focus on improving complex interference and limited wireless bandwidth. Major schemes include the utilisation of non-overlapping channels to avoid interference from nearby nodes (e.g., [9]), the adjusting of transmission rates to avoid interference (e.g., [11]), the scheduling of transmission flows to admit more multicasts (e.g., [12]), the use of licensed RF bands to gain additional bandwidth (e.g., [13]), the integration with wired links to support wireless multicasts using wired transmission capacity (e.g., [2]), etc.

As for wireless multi-hop multicasts with mobile members, studies mostly concern the reliable maintenance of high performance for multicasts when group members change their locations. Researchers have developed and enhanced tree- or mesh-based multicasting protocols to reduce interruptions or interference. The tree multicast in [15] assigns an ID to each multicasting node. The relationship between IDs and the tree root is that the IDs increase in numerical value as they radiate from the root. When supporting multicasts to adapt to changes in link connectivity, the IDs help to confine message exchanges (for repairing a link) to the region where a link breakage occurs. The protocol in [1] builds a shared multicast mesh to maintain group connectivity when members move frequently. It reverses the shortest unicast paths to form multicasting paths on the shared mesh. The rich connectivity of a mesh topology ensures that a new unicast path may always be established when an existing one is interrupted (say due to members' movement). In [10], network function virtualisation is utilised to efficiently share resources when establishing multicasting trees for short delays at mobile edge clouds. With these existing studies, when members move, additional message exchanges are necessary in order to maintain the connectivity of multicasts. Also, these studies do not always include strategies to seamlessly handover mobile members.

\vspace{0.1in}

Geographic multicasting (e.g.,[3]) mitigates the drawback of requiring additional processes and messages on multicasting systems by avoiding frequent multicasting architecture repairs. This is achieved by dividing group members into different zones (say based on members' physical locations) and establishing multicast architectures to connect only zones instead of individual group members. More specifically, for example, in [4], a Steiner tree is computed for zone connections; in [5], a virtual-zone-based structure is proposed to connect zones via a bidirectional tree. Within each zone, broadcasting or greedy forwarding are employed by geographic multicasting to distribute data to all members belonging to a zone. In such a multicast, members' joining or leaving do not cause a zone to move and hence do not require multicasting architectures to be repaired. However, broadcasting unnecessarily consumes the energy of multicasting nodes, presenting a disadvantage for geographic multicasting to be used by UAVs. Moreover, when members move between zones, additional message exchanges may still be needed in order to smoothly handover mobile members. The formation of zones also increases the complexity in establishing a multicasting architecture.

\vspace{0.1in}

Recently, UAV-related wireless multicasts have been popularly studied but have mainly focused on air-to-ground data distribution (e.g., [7-8]). Research issues such as optimal UAV positioning, efficient UAV trajectories, multicarrier modulations, etc. are studied in order to improve the data quality that UAVs can distribute to a group of ground devices. The work in [6] studies reliable and scalable multicasting when UAVs in the multicast move in an ad-hoc manner. An experimental study is carried out to evaluate application-layer rate-adaptive multicast video streaming in an aerial ad-hoc network [14]. In [23], a UAV multicasting scheme is designed to implement multicasting in wireless networks without relying on the underlaying MAC standards. The scheme divides UAVs into clusters so that a UAV with lost data may request retransmissions from a nearby UAV instead of via underlaying MAC messages (e.g., ACKs). In general, these studies investigate a communication scenario different from the one that will be studied in this paper - UAV transitions in aerial multicasts. In [17], we propose a transition algorithm that establishes trajectories in a multicast environment. This study has several drawbacks, e.g., overlooking the seamless possibility of long-distance SLTs, establishing trajectories that are seamless under certain conditions, etc. This paper will not only address these limitations but also improve the study in [17] in terms of searching for a distance-controlled seamless trajectory with controlled computation complexity.

\section{System Model and Problem Formulation}
\subsection{System Model}
\begin{figure}[h]
\begin{center}
\begin{tabular}{c}
\includegraphics[trim=120 465 300 30,clip,height=2.2in]{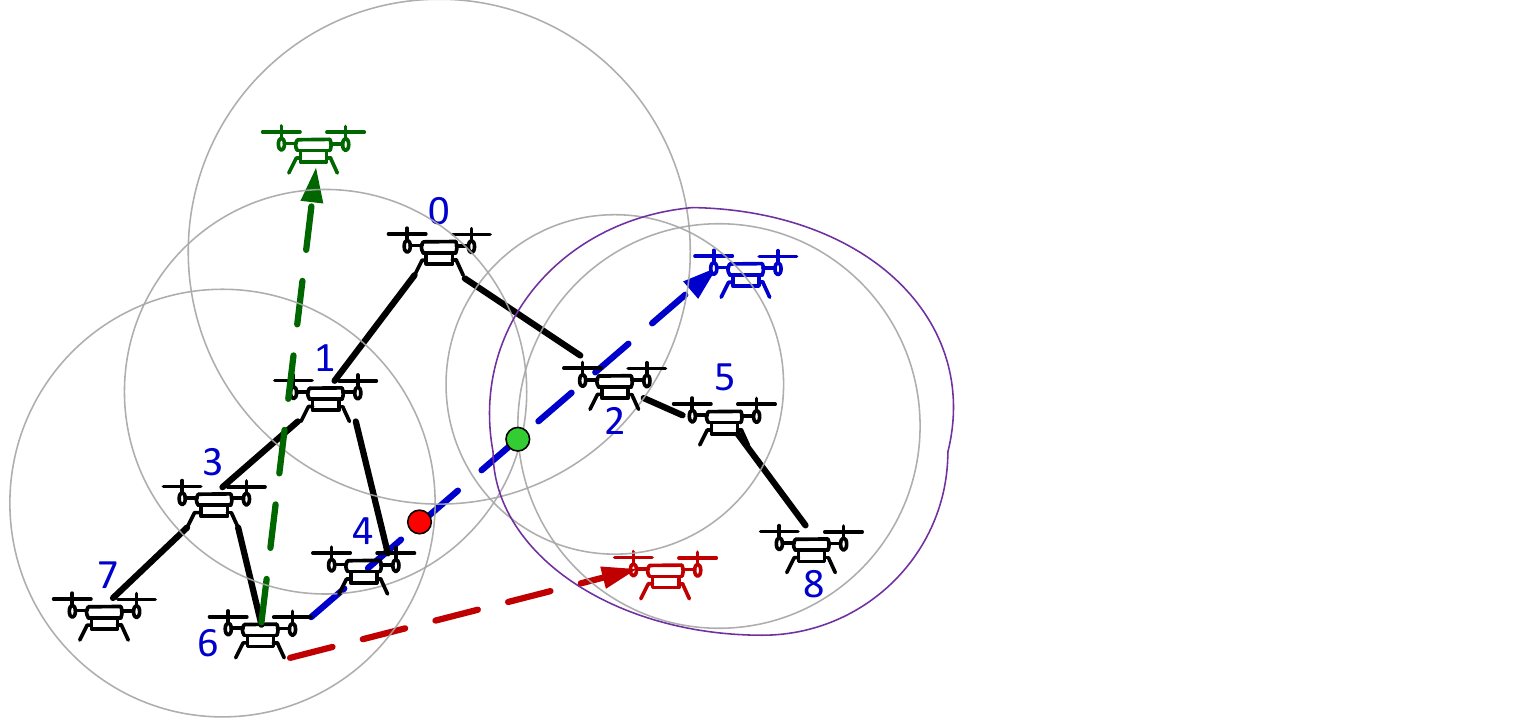}
\end{tabular}
\end{center}
\caption{An example of multi-hop UAV multicasting.}\label{LCRTtree}
\end{figure}
Denote the natural numbers as $\bf{N}$. Suppose there is a group of $n$ ($n\in{\bf{N}}, n>1$) UAVs in our system that are equipped with omnidirectional antennas. Denote the $i$th UAV as $u_i$ ($i\in{[0,n-1]}$) and a mobile UAV as $m$. Let $A$ and $B$ be $m$'s origin and destination, and $F_A$ and $F_B$ be the multicasting forwarders covering $A$ and $B$ respectively. Table I lists the major parameters used in the paper.
\begin{table}[h]
\begin{center}
\caption{List of Major Symbols} \label{parameters1}
\begin{tabular}{|l|l|}
\hline {\bf Symbols} & {\bf Explanations} \\
\hline $n$ & The number of UAVs in the multicasting system. \\
\hline $u_i$ & The $i$th ($i\in{[0,n-1]}$) UAV in the multicasting system. \\
\hline $r_i$ & The radius of referred coverage of $u_i$. \\
\hline $RTR_i$ & The referred coverage/transmission range of $u_i$. \\
\hline $m$ & The mobile UAV. \\
\hline $A$ & $m$'s origin. \\
\hline $B$ & $m$'s destination. \\
\hline $F_A$ & The multicasting forwarder whose coverage covers $A$. \\
\hline $F_B$ & The multicasting forwarder whose coverage covers $B$. \\
\hline $RTR_{F_A}$ & The referred transmission range of $F_A$. \\
\hline $RTR_{F_B}$ & The referred transmission range of $F_B$. \\
\hline $A\rightarrow{B}$ & The straight-line trajectory from $A$ to $B$. \\
\hline $C$ & The intersection between $A\rightarrow{B}$ and the edge of $RTR_{F_A}$ \\
\hline $D$ & The intersection between $A\rightarrow{B}$ and the edge of $RTR_{F_B}$ \\
\hline $T$ & The turning location derived by the efficient trajectory \\
 & formation in Section IV A 2. \\
\hline
\end{tabular}
\end{center}
\end{table}

The signals of a UAV are subject to multi-path reflections from the ground and other potential objects. Such reflections interfere with the UAV's signals differently along different directions, causing the UAV to have an irregular transmission range. Irregular transmission ranges make it complicated to establish a multicast architecture or form paths for mobility. To simplify these processes in order to reduce delays from multicasting and transitions, we employ the referred transmission ranges (RTRs) of UAVs to establish the multicasting architecture and transition paths. Namely, in our system, the link $u_i\rightarrow{u_j}$ can be potentially selected as a multicasting link only if $u_j$ is within the RTR of $u_i$. The referred transmission range of $u_i$ is defined as the spheral space centered at $u_i$ with radius $r_i$. $r_i$ is called the referred coverage radius of $u_i$ and it is the shortest distance of all direct distances from $u_i$ to the edge of $u_i$'s transmission range\footnote{Studies have developed formulas for wireless coverage functions in shadowed or multi-path environment for wireless sensor networks [25-26] or UAV communications [20-21]. $r_i$ can be obtained using these tools.}. Fig.~\ref{LCRTtree} uses grey circles to delimit the RTRs of all UAVs in the example, while the transmission ranges of UAVs are irregular such as the space delimited by the purple closed curve in the figure. Assume that the $n$ UAVs in our system form a multicasting architecture according to an existing wireless multicasting algorithm (e.g., [2]), and referring to UAVs' RTRs.

\subsection{Problem Formulation}
As mentioned, our basic idea to achieve seamless yet resource-efficient UAV transitions is to make use of the combined coverage of multiple forwarders. More specifically, by the combined coverage, we mean the combined RTRs as explained above. Without loss of generality, we refer to two forwarders whose RTRs overlap as overlapping forwarders, and two forwarders whose RTRs do not overlap as non-overlapping forwarders. A simple way to decide whether two forwarders are overlapping or not, is to calculate the Euclidean distance between them. For any two forwarders $u_i$ and $u_j$, if their Euclidean distance is shorter than $(r_i+r_j)$, the two forwarders are overlapping; otherwise, $u_i$ and $u_j$ are non-overlapping. When a UAV transition has origin and destination covered by two overlapping forwarders, we call it a short-distance transition; when a UAV transition has origin and destination covered by two non-overlapping forwarders, we call it a long-distance transition. In Fig.~\ref{LCRTtree}, the transition with the origin and destination connected by the green dotted arrow line is a short-distance transition, and the transition with the origins and destinations connected by the red and the blue dotted arrow lines are long-distance transitions. 

When handling short-distance transitions, in line with our anticipated performance discussed in the Introduction, the determination of the seamlessness of an SLT should be achieved in a resource-efficient manner and without incurring interference to the multicasting system. Our strategy is to enable UAVs to determine whether their SLTs are seamless or not on their own. This will need to equip UAVs with the necessary knowledge of the multicasting topology as well as the intelligence in using the knowledge to determine the seamlessness of SLTs. Similarly, if SLTs are not seamless, our study aims to enable UAVs to establish seamless trajectories on their own. Moreover, these new seamless trajectories also need to offer UAVs with short travel distances in low computation complexity, given that flying is a major energy consumption activity of UAVs and complex computations may delay UAVs' transitions. Therefore, the spontaneous establishment of short seamless trajectories via simple processes forms another of our objectives for short-distance transitions.

When handling a long-distance transition, as with a short-distance transition, we intend to enable a UAV to determine the seamlessness of its SLT. However, the process is more complicated because multicasting forwarders other than the two covering the origin and destination need to be referred to in order to support the check. As shown by the blue dotted arrow lines in Fig.~\ref{LCRTtree}, the different parts of an SLT trajectory are covered by different UAVs and some parts are even covered by more than one UAV, making it complicated to decide which forwarders should be employed together for the seamlessness check. In Fig.~\ref{LCRTtree}, each of the two sets of multicasting UAVs $\{3, 1, 0, 2\}$ and $\{3, 1, 2\}$ 
can be employed to check whether the blue trajectory is seamless or not. The use of $\{3, 1, 2\}$ however yields a faster seamlessness evaluation process. We will hence study how to refer to the minimum number of multicasting forwarders whose RTRs can be employed to carry out fast and resource-efficient seamlessness checks. If an SLT is interrupted, like short-distance transitions, our objective is to enable UAVs to spontaneously establish a new trajectory with a greatly controlled seamless travel distance passing through the combined RTRs of multiple multicasting forwarders. In the following sections, we present our solutions for these research problems.

\section{The Efficient Transition via Trajectory Adjustment}
This section studies efficient and seamless short-distance transitions (Section IV. A) and long-distance transitions (Section IV.B). We then systematically combine these results to present the efficient transition formation (ETF) algorithm (Section IV.C) for resource-efficient and seamless UAV transitions in a multicast environment.

\subsection{Seamless Short-distance UAV Transitions}
This subsection studies UAV transitions between overlapping forwarders, by first analysing the conditions for an SLT to be seamless and then forming an efficient trajectory to replace an interrupted SLT with a new short seamless trajectory.

\subsubsection{The Seamlessness Condition of A Straight-line Trajectory for Short-distance Transitions}
\begin{figure}[h]
\begin{center}
\begin{tabular}{c}
\includegraphics[trim=80 400 90 130,clip,height=2.2in]{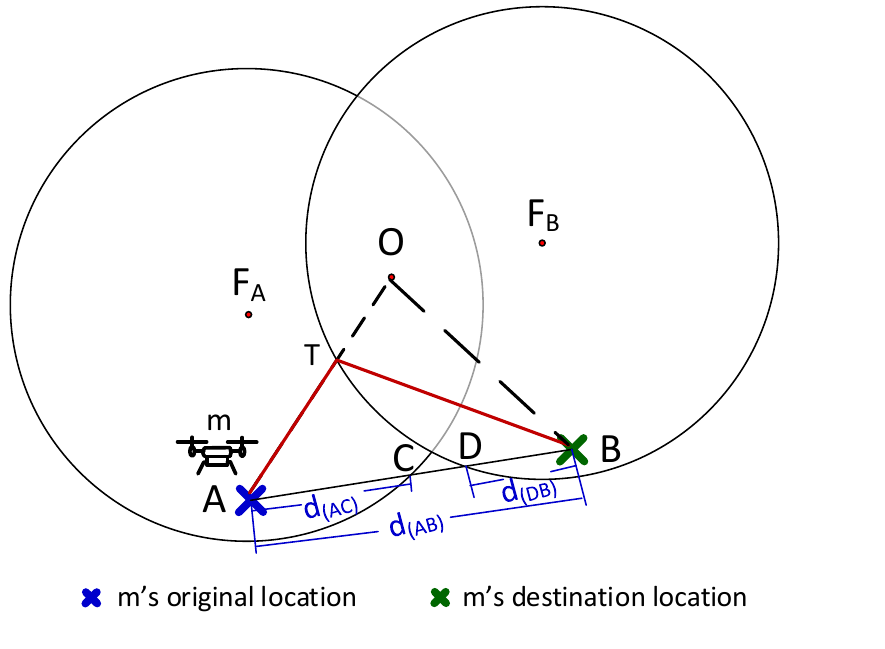}
\\
\end{tabular}
\end{center}
\caption{Deriving the trajectory adjustment condition.}\label{tac}
\end{figure}
We use Fig.~\ref{tac} to illustrate our idea to determine whether a short-distance SLT is seamless or not. In this figure, $m$ moves from $A$ to $B$, and $F_A$ and $F_B$ are the forwarders covering $A$ and $B$. Let the intersections between the SLT $A\rightarrow{B}$ and the RTR edges of $F_A$ and $F_B$ be $C$ and $D$ respectively. Denote the distances between $A$ and $B$, $A$ and $C$, and $B$ and $D$ as $d_{(AB)}$, $d_{(AC)}$, and $d_{(BD)}$. Theorem 1 gives the condition that $m$ may transit via the SLT $A\rightarrow{B}$.

{\bf Theorem 1.} {\it For a UAV $m$ moving between locations covered by two overlapping forwarders (e.g., in Fig.~\ref{tac}) on a multicasting architecture, the SLT connecting $m$'s origin and destination is seamless if one of the following two conditions meets: \\
1) $d_{(AB)}\leq{d_{(AC)}+d_{(DB)}}$, or \\
2) when $d_{(AB)}>{d_{(AC)}+d_{(DB)}}$, there exists a forwarder $u_i$ in the multicasting system whose distances to $C$ and $D$ are both $\leq{r_i}$, where $r_i$ is the referred coverage radius of $u_i$.} 

{\it Proof.} We prove Theorem 1 by contradiction. When $d_{(AB)}\leq{d_{(AC)}+d_{(DB)}}$, suppose SLT is not seamless. Then, one or more part(s) of the SLT are not covered by $F_A$ or $F_B$. Let the length of the uncovered trajectory part(s) be $l>0$. We then have
$
d_{(AC)}+l+d_{(DB)}=d_{(AB)}\Rightarrow{d_{(AB)}>{d_{(AC)}+d_{(DB)}}}.
$
This contradicts $d_{(AB)}\leq{d_{(AC)}+d_{(DB)}}$. Therefore, when $d_{(AB)}\leq{d_{(AC)}+d_{(DB)}}$, the SLT is seamless.

When $d_{(AB)}>{d_{(AC)}+d_{(DB)}}$, and there exists a multicasting forwarder $u_i$ whose distances to $C$ and $D$ are both $\leq{r_i}$, where $r_i$ is $u_i$'s referred coverage radius, suppose the SLT is not seamless. Then, there is at least a point between $C$ and $D$ whose Euclidean distance to this forwarder is $>r_i$. This causes $C\rightarrow{D}$ not to be a straight line because the two ends $C$ and $D$ are both within the distance of $r_i$ to the forwarder. This contradicts the fact that $A\rightarrow{B}$ is a straight line. Q.E.D.

The implementation of Theorem 1 requires knowledge of the Euclidean coordinates of $C$ and $D$: $(x_C, y_C, z_C)$ and $(x_D, y_D, z_D)$.
In order to obtain $(x_C, y_C, z_C)$, as $C$ is on the edge of $F_A$'s RTR (i.e., $RTR_{F_A}$), we have
\begin{equation}
(x_C-x_{F_A})^2+(y_C-y_{F_A})^2+(z_C-z_{F_A})^2=r_{(F_A)}^2.\label{theorem1Equ1}
\end{equation}
Also, $C$ is on $A\rightarrow{B}$, meaning
\begin{equation}
\begin{cases}\label{theorem1Equ2}
x_C=x_A+t(x_B-x_A), \\
y_C=y_A+t(y_B-y_A), \\
z_C=z_A+t(z_B-z_A).
\end{cases}
\end{equation}
Combining (\ref{theorem1Equ1}) and (\ref{theorem1Equ2}), $m$ can obtain $t$ by solving the quadratic equation for $t$. Typically two distinct values of $t$ exist, defining two distinct points. The point which is closer to $F_B$ is $C$. Similarly, $m$ obtains $D$'s coordinates, and then calculates $d_{(AB)}$, $d_{(AC)}$, and $d_{(DB)}$ for Theorem 1. Therefore, Theorem 1 can be implemented by using the 3-dimensional Pythagorean formula, yielding a fast yet resource-efficient process. As for seeking the forwarders that may cover both $C$ and $D$, the process can be limited to those forwarders that overlap with $F_A$ and $F_B$, controlling the time complexity of Theorem 1. While Theorem 1 is simple and efficient, it may not provide a full understanding regarding the seamlessness of an SLT. For example, an SLT may be fully covered by more than one multicasting forwarder. Our efficient seamlessness checking algorithm designed in Section IV.B.1 can be employed to implement a thorough seamlessness check, i.e., to find whether the RTRs of multiple forwarders can together cover an SLT seamlessly. Whether such a thorough seamlessness check is necessary for a short-distance transition depends on the application requirements and user preferences, as the difference in travel distances following an SLT or a new trajectory developed in the next subsection may not be significant. Also, in a multicasting architecture built up to avoid interference between forwarders, the chance that multiple forwarders overlap with $F_A$ and $F_B$ is greatly constrained. Therefore, in this paper, if the conditions in Theorem 1 are not met, $m$ will establish a new trajectory to transit as in the next subsection.

\subsubsection{The Efficient Trajectory Formation for Short-distance Transitions}
When proposing a new trajectory, we try to control $m$'s travel distances as well as extra traffic overheads with low complexity processing, allowing fast transitions with efficient use of resources (e.g., energy, bandwidth). The idea is to employ a location (denoted as $T$), within the overlap of $RTR_{F_A}$ and $RTR_{F_B}$, to form a transition path $A\rightarrow{T}\rightarrow{B}$ inside the combined RTR of $F_A$ and $F_B$. We refer to $T$ as the turning location of the new trajectory. Ideally, $T$ should minimise the extra travel distance exceeding that of $m$'s SLT. Such a location is achievable by an existing algorithm (e.g., [27]) to seek a point on the surface of the overlapping area that has the shortest distance to $A\rightarrow{B}$. However, the complex computation potentially delays $m$'s transitions which affects $m$'s ability to meet real-time performance requirements. We hence propose a heuristic turning location to balance the tradeoff between the low computation complexity and short travel distance in finding a seamless trajectory.

Our basic idea is to employ the intersection between $RTR_{F_B}$ and the straight line $O\rightarrow{A}$ as $T$, where $O$ is the center of the overlap of $RTR_{F_A}$ and $RTR_{F_B}$. This is because the location of $O$ can be obtained as $(\frac{x_A+x_B}{2},\frac{y_A+y_B}{2},\frac{z_A+z_B}{2})$, requiring much less computation than any other point in the 3-dimensional overlapping area. The reason that $m$ does not use $O$ directly is because $T$ on the overlap surface always has a shorter overall distance to $A$ and $B$ than $O$. This can be proved by referring to Fig.~\ref{tac}. We use $|AT|$, $|BT|$, $|AO|$, $|OT|$, and $|BO|$ to represent the lengths of the line segments $AT$, $BT$, $AO$, $OT$, and $BO$. Obviously, $|AT|<|AO|$. Then, based on the Triangle Inequality Theorem, for the triangle $OTB$, $|BT|<|OT|+|BO|$. Therefore, we have $|AT|+|BT|<|AT|+|OT|+|BO|=|AO|+|BO|$. Since $T$ is covered by both $F_A$ and $F_B$, the lines $A\rightarrow{T}$ and $T\rightarrow{B}$ can be covered by $F_A$ and $F_B$, ensuring the seamlessness of $A\rightarrow{T}\rightarrow{B}$.

By the line function between $A$ and $O$, $T$'s coordinates can be formulised as below,
\begin{equation}\label{lineequation}
\begin{cases}
x_T = t(\frac{x_A+x_B}{2}-x_A)+x_A, \\
y_T = t(\frac{y_A+y_B}{2}-y_A)+y_A, \\
z_T = t(\frac{z_A+z_B}{2}-z_A)+z_A. \\
\end{cases}
\end{equation}
Furthermore, as $T$ is on the edge of $F_B$'s coverage, we have $(x_T-x_{F_B})^2+(y_T-y_{F_B})^2+(z_T-z_{F_B})^2=r^2$. Combining this equation with (3), we can derive $t$ and hence $T$'s coordinates. The new trajectory $A\rightarrow{T}\rightarrow{B}$ is shown by the red lines in Fig.~\ref{tac}.

\subsection{Seamless Long-distance UAV Transitions}
\subsubsection{The Efficient Seamlessness Checking Algorithm for Long-distance SLTs}
When $m$ transits a long distance, as analysed in Section III.B, multiple sending/forwarding UAVs need to be selected in order to decide the seamlessness of $m$'s SLT. Obviously, each selected UAV should cover part of $A\rightarrow{B}$. Namely, a selected UAV should have a distance to $A\rightarrow{B}$ less than its referred coverage radius. Without loss of generality, we refer to such selected UAVs as the covering forwarders of $A\rightarrow{B}$. In order to find these covering forwarders, $m$ needs to calculate the distance of each multicasting forwarder to $A\rightarrow{B}$. Suppose there are $n_f$ multicasting forwarders in the system. For each of these forwarders $u_i (i\in{[0,n_f-1]})$, based on Heron's formula, its distance to $A\rightarrow{B}$ can be calculated by
\begin{equation}
\dot{d}=\frac{2\sqrt{L(L-|AB|)(L-|A{u_i}|)(L-|{u_i}B|)}}{|AB|},\label{coveringforwarer}
\end{equation}
where $|AB|$, $|A{u_i}|$, and $|{u_i}B|$ are the Euclidean distances between $A$ and $B$, $A$ and $u_i$, and $u_i$ and $B$ respectively, and $L=\frac{|AB|+|A{u_i}|+|{u_i}B|}{2}$. $u_i$ is a covering forwarder of $A\rightarrow{B}$ if $\dot{d}\leq{r_i}$, where $r_i$ is the referred coverage radius of $u_i$. Equation (\ref{coveringforwarer}) requires $m$ to know the coordinates of multicasting forwarders. Recall, $m$ obtains such information from $F_A$ while establishing the multicasting architecture.

\vspace{0.1in}

If $F_A$ and $F_B$ are the only covering forwarders found by $m$ for $A\rightarrow{B}$, the part of $A\rightarrow{B}$ that is outside of the RTRs of $F_A$ and $F_B$ cannot be covered by other multicasting forwarders. Therefore, $A\rightarrow{B}$ is not seamless in this case. But if $m$ finds more covering forwarders, further processing is required to determine the seamlessness of $A\rightarrow{B}$. In general, our idea is to progressively determine checking points, on $A\rightarrow{B}$, toward $B$ and check whether these checking points may be covered by at least one covering forwarder or not. Checking points are those points at which $A\rightarrow{B}$ leaves the RTRs of certain covering forwarders. Since a checking point is where $A\rightarrow{B}$ leaves the coverage of a current covering forwarder, if $m$ cannot find another covering forwarder to cover this point, $A\rightarrow{B}$ cannot be seamless and $m$ terminates the checking process. Otherwise, if the process reaches a checking point that can be covered by $F_B$, $A\rightarrow{B}$ is seamless. In order for $m$ to implement a fast, resource-efficient yet comprehensive seamlessness checking process, our algorithm enables $m$ to select a minimal number of checking points. This is achieved by prioritising potential points that have shorter Euclidean distances to $B$ as checking points. The detailed procedure is described below.

There are two intersections between the edge of $RTR_{F_A}$ and $A\rightarrow{B}$. The one closer to $B$ is set as the first checking point by $m$. This first checking point is denoted as $I_0$. Also, $m$ initialises a checked forwarder list $CL$. $CL$ records the forwarders that we have so far selected to cover checking points, and to choose successive checking points. For example, as $I_0$ is where $A\rightarrow{B}$ leaves the RTR of $F_A$, $m$ adds $F_A$ to $CL$. Then, from all covering forwarders that are not in $CL$, $m$ selects those that can cover $I_0$. If there is no such covering forwarder, $m$ decides that $A\rightarrow{B}$ is not seamless. But if there is at least one such covering forwarder, for each of them, $m$ calculates the intersections between its RTR and $A\rightarrow{B}$. Among these calculated intersections, $m$ chooses the one closer to $B$ as the second checking point $I_1$ and adds the forwarder for which it was calculated to $CL$. $m$ checks whether $I_1$ is in the RTR of $F_B$ (the forwarder covering $m$'s destination) by calculating their Euclidean distance. If yes, $A\rightarrow{B}$ is seamless. Then, just as for $I_0$, $m$ checks whether $I_1$ can be covered by at least one covering forwarder. The procedure continues until $m$ finds 1) a checking point that cannot be covered by the RTRs of any covering forwarders, showing that $A\rightarrow{B}$ is not seamless, or 2) a checking point that can be covered by the RTR of $F_B$, meaning that $A\rightarrow{B}$ is seamless. Algorithm 1 systematically presents the detailed steps of our efficient seamlessness checking algorithm.
\begin{tabbing}
\renewcommand{\baselinestretch}{1}
---------------------------------------------------------------------------\\[-6pt]
{\bf\small Algorithm 1 The Efficient Seamlessness Checking Algorithm} \\ {\bf\small for Long-distance SLTs}
\\ xxxxxx\=xxx\=xxx\=xxx\=xxx\=xxx\=xxx\=xxx\= \kill \small Input:
\> Transition UAV $m$, $A$ and $B$ (i.e., $m$'s origin and \\ destination), $F_A$ and $F_B$ (i.e., forwarders covering $A$ and $B$); \\
Output: \> $m$'s decision on the seamlessness of $A\rightarrow{B}$. \\
---------------------------------------------------------------------------\\[-6pt]
1. $m$ calculates the distance from each forwarder to $A\rightarrow{B}$ \\ by (\ref{coveringforwarer}), and selects covering forwarders; \\
2. If $F_A$ and $F_B$ are the only covering forwarders, $A\rightarrow{B}$ \\ is not seamless; $m$ exits. \\ 
3. Otherwise, \\
4. \> $m$ initialises $i=0$ and $CL$ as empty; \\
5. \> $m$ calculates the checking point $I_i$ based on the RTR \\ of $F_A$ and $A\rightarrow{B}$; \\ 
6. \> If $I_i$ can be covered by $F_B$, $A\rightarrow{B}$ is seamless; \\ $m$ exits. \\
7. \> If $I_i$ cannot be covered by $F_B$ \\
8. \>\> $m$ adds to $CL$ the forwarder for which $I_i$ is the \\ point where $A\rightarrow{B}$ leaves its RTR; \\
9. \>\> Among the covering forwarders not in $CL$, $m$ \\ seeks those whose RTRs can cover $I_i$; \\
10. \>\> If there is no such forwarder, $m$ decides that \\ $A\rightarrow{B}$ is not seamless; $m$ exists. \\
11. \>\> If there are forwarders selected, \\
12. \>\>\> $i=i+1$; \\
13. \>\>\> For each of these forwarders, $m$ calculates the \\ intersections of their RTRs with $A\rightarrow{B}$; \\
14. \>\>\> Among these intersections, $m$ calculates their \\ Euclidean distances to $B$; $m$ selects the intersection with \\ the shortest distance to $B$ as $I_i$ and goes to Step 6; \\
---------------------------------------------------------------------------\\[-6pt]
\end{tabbing}

Algorithm 1 determines the seamlessness of $A\rightarrow{B}$ by preferentially checking the points that are closer to the destination, benefitting a fast checking process. The complexity of Algorithm 1 is $O(n'n_f)$, where $n_f$ is the number of forwarders in the multicasting system and $n'$ is the number of covering forwarders of $A\rightarrow{B}$. We use the example in Fig.~\ref{LCRTtree} to illustrate Algorithm 1. UAV 6 is transiting and the blue dotted arrow line shows its SLT. By Step 1, UAV 6 finds that UAVs 3, 1, 0, 2, and 5 are covering forwarders. Then, by Step 5, $m$ select the intersection indicated by the red dot as $I_0$. As $I_0\neq{F_B}$, UAV 6 seeks all forwarders that can cover $I_0$ (Step 9). UAV 1 is the only forwarder that UAV 6 selects. Based on Steps 11 - 14, UAV 6 sets the intersection (indicated by the green dot in the figure) as $I_1$. Then, via Step 6, since $I_1$ is within the RTR of UAV 5 (the $F_B$ in this example), UAV 6 decides that the blue dotted arrow line is seamless and exits the algorithm.

\subsubsection{The Efficient Seamless Trajectory Formation Algorithm for Long-distance Transitions}
When proposing a seamless trajectory for a long-distance transition, our strategy is to search for a chain of multicasting forwarders, starting from $F_A$ and ending at $F_B$, whose RTRs overlap one by one so that their combined RTRs provide an opportunity for $m$ to travel seamlessly. For the sake of fast transitions, this forwarder chain should consist of the minimal number of multicasting forwarders. Also, in order to minimise the delays before $m$ can start transiting, we will obtain such a forwarder chain with controlled computational complexity. To achieve this, our algorithm processes differently for long-distance transitions with origin and destination RTRs on the same multicasting paths and long-distance transitions with origin and destination RTRs on different multicasting paths.

In detail, recall that in our system the information of locations and referred coverage radii of multicasting forwarders piggybacks on the control packets issued while establishing a multicasting architecture. $m$ receives this information from $F_A$ when it joins the multicasting architecture. By referring to this information, $m$ is able to find out whether its transition is between the RTRs of two forwarders that are on the same multicasting path or not. If yes, the multicasting forwarders on the path between $F_A$ and $F_B$ naturally forms the forwarder chain, without requiring further computing tasks at $m$. If no, $m$ searches for the minimum number of multicasting forwarders whose RTRs can be combined to offer a seamless trajectory for $m$. More specifically, starting from $F_A$, $m$ employs Dijkstra's algorithm\footnote{Other algorithms such as the A$^{\ast}$ search algorithm are also suitable for our purpose.} to seek these forwarders on the chain. Since $m$ intends to search for a trajectory with a short travel distance so as to achieve an energy-efficient transition, it employs the Euclidean distances between multicasting forwarders as the weights of links between forwarders when computing the forwarder chain by Dijkstra's algorithm.

With the selected forwarder chain, $m$ starts establishing the new trajectory through the combined RTRs of forwarders on the chain. This new trajectory is formed by a set of straight lines that connect a group of turning locations achieved by the efficient trajectory formation scheme in Section IV.A.2. More specifically, suppose there are $n''$ forwarders on the chain. Denote $F_A$ as $F_0$, $F_B$ as $F_{(n''-1)}$, and other forwarders on the chain as $F_i (i\in{[1,n''-2]})$. $m$ initiates the first turning location $T_0$ as $m$'s origin $A$. Then, for each pair of forwarders $F_j$ and $F_{(j+1)}$ ($j\in{[0,n''-2]}$), $m$ seeks the intersection between the line segment $T_{j}\rightarrow{O_{j,(j+1)}}$ and the RTR of $F_{(j+1)}$, where $O_{j,(j+1)}$ is the center of overlap of RTRs of $F_{j}$ and $F_{(j+1)}$. This intersection is set as the $(j+1)$th turning location $T_{(j+1)}$. Once all $(n''-1)$ turning locations are found, the new trajectory is formed by connecting these turning locations one by one via straight lines, i.e., $A\rightarrow{T_1}\rightarrow{T_2}\rightarrow{...}\rightarrow{T_{(n''-1)}}\rightarrow{B}$. Algorithm 2 systematically presents the above processes.

\begin{tabbing}
\renewcommand{\baselinestretch}{1}
---------------------------------------------------------------------------\\[-6pt]
{\bf\small Algorithm 2 The Efficient Seamless Trajectory Formation} \\ {\bf\small Algorithm for Long-distance Transitions}
\\ xxxxxx\=xxx\=xxx\=xxx\=xxx\=xxx\=xxx\=xxx\= \kill \small Input:
\> Transition UAV $m$, $A$ and $B$, $F_A$ and $F_B$; \\
Output: \> $m$'s shortest seamless trajectory. \\
---------------------------------------------------------------------------\\[-6pt]
1. $m$ determines whether $A$ and $B$ are covered by multicast- \\ ing forwarders on the same multicasting path; \\
2. If yes, $F_A$, $F_B$, and the multicasting forwarders between \\ $F_A$ and $F_B$ on the path form the forwarder chain; \\
3. If no, $m$ employs Dijkstra's algorithm and uses the dista- \\ nces between multicasting forwarders as link weights to \\ search for the minimum number of forwarders between $F_A$ \\ and $F_B$; These selected forwarders with $F_A$ and $F_B$ form \\ the forwarder chain; \\
4. $m$ sets $i=0$, $F_0=F_A$, other forwarders on the chain \\ as $F_i (i\in{[1,n''-2]})$, and $T_0=A$; \\
5. $j=0$; \\
6. While $j<(n''-2)$ \\
7. \> $m$ calculates the intersection between the RTR of \\ $F_{(j+1)}$ and the line segment $T_j\rightarrow{O_{j,(j+1)}}$; $m$ sets this \\ intersection as the $(j+1)$th turning location $T_{(j+1)}$; $j=$ \\ $j+1$; \\
8. $m$ connects the selected turning locations one by one \\ to form its seamless trajectory; $m$ exits. \\
---------------------------------------------------------------------------\\[-6pt]
\end{tabbing}
\begin{figure}[h]
\begin{center}
\begin{tabular}{c}
\includegraphics[trim=120 480 10 30,clip,height=2.0in]{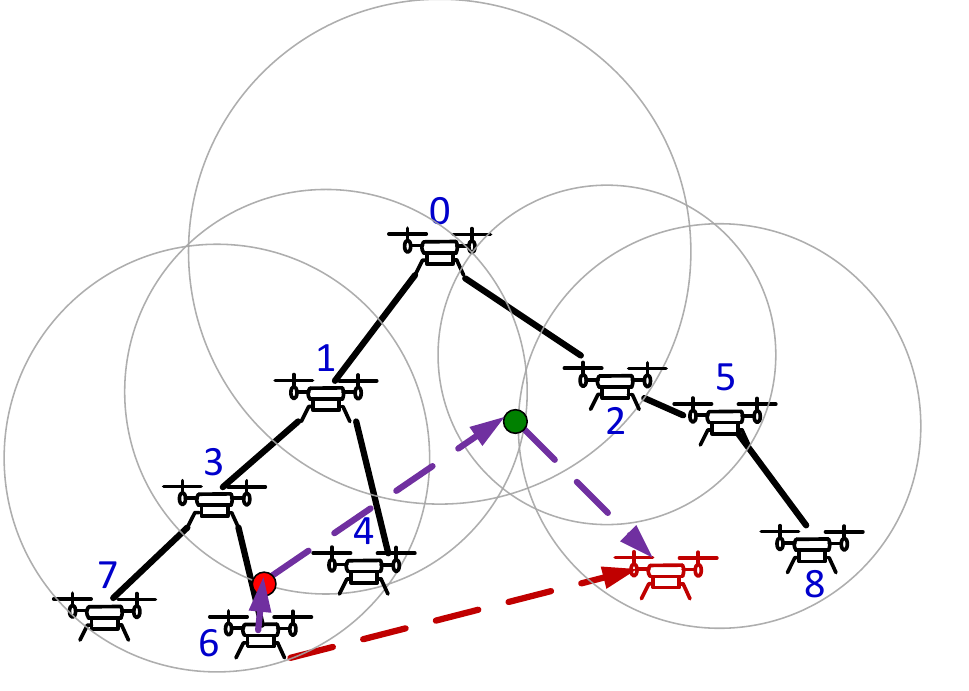}
\end{tabular}
\end{center}
\caption{An example of forming the shortest seamless trajectory for a long-distance transition.}\label{Algorithm2}
\end{figure}

In the example of Fig.~\ref{Algorithm2}, the red dotted arrow line is not seamless. As the origin and the destination are not covered by forwarders on the same path, by Step 3, UAV 6 forms the forwarder chain with the minimum number of forwarders based on their Euclidean distances. UAVs 3, 1, 5 will be selected in order. In Step 4, UAV 6 sets $F_0$ as UAV 3, $F_1$ as UAV 1, and $F_2$ as UAV 5. UAV 6 sets $j=0$. By Step 7, UAV 6 determines $T_1$ which is illustrated as the red dot in the figure. Then, by the same step, UAV 6 determines $T_2$ which is illustrated as the green dot in the figure. At this time, as $j=1$ and $n''=3$, UAV 6 goes to Step 8 to form the new trajectory $A\rightarrow{T_1}\rightarrow{T_2}\rightarrow{B}$, as illustrated as the purple dotted arrow lines in the figure.

\subsection{The ETF Algorithm}
The analyses and algorithms studied in Sections IV.A and IV.B can be easily integrated with an existing multi-hop multicasting algorithm, because the implementation of these results only require knowledge of the coordinates of multicasting forwarders. As mentioned, coordinates are obtainable for example via a GPS receiver or as discussed in previous studies (e.g., [18-19]), and can be piggybacked on the control packets exchanged between multicasting forwarders while establishing the multicasting architecture. In this section, we combine these results so that $m$ may efficiently choose an appropriate trajectory for the transition $A\rightarrow{B}$ under a variety of different conditions.
\begin{tabbing}
\renewcommand{\baselinestretch}{1}
---------------------------------------------------------------------------\\[-6pt]
{\bf\small Algorithm 3 Efficient Trajectory Formation Algorithm}
\\ xxxxxx\=xxx\=xxx\=xxx\=xxx\=xxx\=xxx\=xxx\= \kill \small Input:
\> The UAV multicasting architecture, the mobile UAV \\ $m$, $A$ and $B$, and $F_A$ and $F_B$; \\
Output: \> $m$'s ETF transition trajectory. \\
---------------------------------------------------------------------------\\[-6pt]
1. When establishing the multicasting architecture, all select- \\ ed
multicasting forwarders exchange the information of their \\ locations and referred coverage radii; these forwarders calcu- \\ late and exchange their Euclidean distances with each other; \\
2. $m$ decides whether $F_A$ and $F_B$ are overlapping or not ba- \\ sed on the Euclidean distance of $F_A$ and $F_B$; \\
3. If overlapping (i.e., a short-distance transition), $m$ employs \\ Theorem 1 to check seamlessness of $A\rightarrow{B}$; \\
4. \> If seamless, $m$ transits via $A\rightarrow{B}$ and exits. \\
5. \> Otherwise, $m$ forms a seamless trajectory by the eff- \\ icient seamless trajectory formation scheme (Section IV.A.2), \\ and exits. \\
6. If non-overlapping (i.e., a long-distance transition), \\
7. \> $m$ employs Algorithm 1 to check the seamlessness \\ of its SLT $A\rightarrow{B}$; \\
8. \> If seamless, $A\rightarrow{B}$ is $m$'s trajectory, and $m$ exits. \\
9. \> Otherwise, $m$ employs Algorithm 2 to form the \\ seamless trajectory, and exits.\\
---------------------------------------------------------------------------\\[-6pt]
\end{tabbing}

The ETF algorithm recruits UAVs on the multicasting architecture to implement seamlessness checks and establish seamless trajectories in a resource-efficient and delay-controlled fashion. Its time complexity can be analysed by looking into short-distance transitions and long-distance transitions separately.
\begin{itemize}
\item With a short-distance transition, the ETF algorithm checks the seamlessness of this SLT (Section IV A 1) by computing the equation system (2) and 3-dimensional Pythagorean formulae for maximally $(n_f-1)$ multicasting forwarders, where $n_f$ is the number of multicasting forwarders in the system. Therefore, this process has the time complexity of $O(n_f)$. The trajectory formation for short-distance transitions (Section IV A 2) takes a constant time because it only calculates the equation system (3) once. Overall, the time complexity of our ETF algorithm in handling a short-distance transition is $O(n_f)$.
\item With a long-distance transition, the ETF algorithm checks the seamlessness of this SLT by Algorithm 1 (Section IV B 1). In order to seek covering forwarders, Algorithm 1 calculates 3-dimensional Pythagorean formulae and the equation (4) for $n_f$ multicasting forwarders, which follows the time complexity of $O(n_f)$. Among the $n'$ selected covering forwarders, by calculating 3-dimensional Pythagorean formulae, Algorithm 1 determines checking points and then seeks those covering forwarders that can cover these checking points, resulting in a time complexity of $O(n')$. Since $n'\leq{n_f}$, Algorithm 1 follows $O(n_f)$. When a long-distance SLT is not seamless, the ETF algorithm employs Algorithm 2 (Section IV B 2) to form a new seamless trajectory. Dijkstra's algorithm is employed to seek the minimum number of transition forwarders among $n_f$ multicasting forwarders, meaning the time complexity is $O({n_f}^2)$. However, this complexity may drop to a constant time for a transition with the origin and destination covered by multicasting forwarders on the same multicasting path. Overall, the time complexity for our ETF algorithm to handle a long-distance transition is $O({n_f}^2)$.
\end{itemize}
To summarise the above analysis, our ETF algorithm follows $O({n_f}^2)$.

The ETF Algorithm forms seamless and efficient trajectories to support the transition of a multicasting receiver or a multicasting sender/forwarder. When a UAV sender/forwarder plans to transit, it may incur additional interference or cause a disconnected multicasting architecture.
\begin{itemize}
\item To mitigate additional interference, many studies have proposed effective solutions such as utilising channel diversity, scheduling transmission nodes, hopping between channels, adjusting transmission coverage, etc. Because UAVs have limited energy resources, the use of multiple channels is more attractive as other schemes require mobile UAVs to implement more complicated processes. Furthermore, in aerial environments, more wireless bands are unoccupied which may be employed by UAVs as potential new transmission resources.
\item To restore interrupted multicasting architectures, existing studies (e.g., [5]) have proposed schemes/algorithms to deal with dynamic group memberships. While the ETF Algorithm can be easily integrated with these existing proposals and adopt their mobility schemes to reconnect the interrupted part(s) to the multicasting architecture, these mobility schemes often cost energy and bandwidth to issue control overheads. Hence, for each mobile UAV sender or forwarder, we also propose the introduction of a new UAV to the system. Given the limited energy supply and the fast consumption of  energy during movement, a UAV that has already served as a sender or a multicasting forwarder may be unable to complete its mobile missions with its remaining energy. We hence nominate the new UAV to move instead of the sender/forwarder, enabling UAV transition activities while maintaining a reliable multicasting architecture. As UAVs are low-cost communication devices, the use of further UAVs for the multicasting system should not cause a great additional financial cost.
\end{itemize}
As for receiving-only UAVs, their movements do not affect data multicasting, i.e., do not cause interference or disconnected architectures. Hence, Algorithm 3, i.e., the ETF Algorithm, can be employed to transit such UAVs without additional actions as above.

\section{Simulation Evaluations}
In this section, by conducting experimental studies with the discrete event network simulator NS2.35 [16], we compare the following five multicasting schemes when they handle mobile group members.
\begin{itemize}
\item LCRT, proposed in [2], does not provide transition support for mobile group members. To establish a LCRT multicasting tree, UAVs are assigned to different levels based on their shortest hop distances to the sending source. At each level, UAVs that can cover more forwarding/receiving UAVs at the immediately higher level are selected as multicasting forwarders with priority (limiting the number of forwarders for reduced interference), until all forwarding/receiving UAVs in the immediately higher level have found their forwarders.
\item T-LCRT that provides transition support for mobile UAVs on a LCRT multicasting tree. Transition forwarders are selected by negotiations between UAVs. A UAV receiver may be asked to forward data to a mobile UAV when this mobile UAV is moving close to it.
\item EGMP. A geographic multicasting that groups UAVs into zones and connects these zones via a bi-directional multicasting tree [5]. More details can be referred to in Section II.
\item ETTA. It employs the LCRT algorithm [2] to establish the multicasting architecture. UAV transitions on the LCRT tree are supported by the ETTA algorithm in [17].
\item ETF. Like ETTA, it establishes a multicasting architecture by the LCRT algorithm [2]. UAV transitions are supported by the ETF algorithm presented in this paper.
\end{itemize}
As our goal for the ETF algorithm is to maintain high-performance multicasting for not only transiting UAVs but also other group receivers, we evaluate the above five schemes along the following metrics in our simulations.
\begin{itemize}
\item Average multicast delay (AMD) of the multicasting group. In our simulations, AMD is calculated by
$$
AMD=\frac{AD_i}{n}, i\in{[0,n-1]},
$$
where $AD_i$ is the average packet delay at the $i$th UAV receiver, and $n$ is the total number of UAV receivers in the group. AMD is observed to evaluate how UAV transitions affect the real-time delivery of multicasting data to all receivers. A short AMD is anticipated, as it means that multicasting communications are still potentially in real time when UAV transitions take place.
\item Average multicast throughput (AMT) of the multicasting group. In our simulations, AMT is calculated by
$$
AMT=\frac{AT_i}{n}, i\in{[0,n-1]},
$$
where $AT_i$ is the average data throughput at the $i$th UAV receiver. AMT is observed to evaluate how UAV transitions affect multicasting interference or disconnections. A high AMT means low multicasting data loss rate and hence the success of UAV transition schemes in controlling interference and disconnections.
\item Average mobile delay (AMoD). This refers to the average delay achieved by mobile UAVs. We calculate AMoD by
$$
AMoD=\frac{MD_j}{n_m}, j\in{[0,n_m-1]},
$$
where $MD_j$ is the average packet delay at the $j$th mobile UAV receiver, and $n_m$ is the total number of mobile UAVs in the multicast. AMoD shows whether a transition UAV may receive data in real time. A short AMoD contributes to good performance for a transition UAV in terms of receiving data in real time.
\item Average mobile throughput (AMoT). This refers to the average throughput received at mobile UAVs. We calculate AMoT by
$$
AMoT=\frac{MT_j}{n_m},j\in{[0,n_m-1]},
$$
where $MT_j$ is the average packet throughput at the $j$th mobile UAV receiver. AMoT helps to demonstrate whether a UAV is disconnected from the multicasting during its transition or not. A high AMoT means a low data loss rate at a transition UAV, indicating the positive effect of the transition schemes. 
\item Average additional energy consumption per received bit (AAEB). For each UAV, we define its additional energy consumption per received bit (AEB) as the ratio of the total amount of energy used to support its transition to the total number of bits it receives. The AEB of the $i$th UAV ($i\in{[0,n_t-1]}$) is calculated by
    $$
    AEB_i=\frac{E_i}{B_i},
    $$
    where $n_t$ is the number of transition UAVs in the system, $E_i$ is the total amount of energy consumed for the $i$th UAV to fly to its destination and for the $i$th UAV and other UAVs to exchange control messages to support the $i$th UAV's transition, and $B_i$ is the number of bits received at the $i$th UAV. Then, AAEB is obtained by
    $$
    AAEB=\frac{\sum_{i=0}^{n_t-1}AEB_i}{n_t}.
    $$
    A lower AAEB means less energy used to support UAV transitions in a multicasting system.
\item Additional control overhead (ACO). ACO refers to the extra control traffic generated by a multicasting scheme in order to support the transition of mobile UAVs. A lower ACO mainly means less bandwidth consumption when supporting UAV transitions in a multicasting communication.
\end{itemize}

We conduct three simulations with common settings listed in Table II. The results plotted in the figures of this section are based on the mean values of 20 simulation runs.
\begin{table}[h]
\begin{center}
\caption{Simulation Parameters} \label{parameters1}\vspace{1em}
\begin{tabular}{|l|l|}
\hline {\bf Parameter} & {\bf Value} \\
\hline Frequency & 2.4GHz \\
\hline UAV altitudes & 40m-100m \\
\hline Propagation model & Free space \\
\hline Dimensions & 3D \\
\hline Transmission power & 15dBm \\
\hline Number of channels & 1 \\
\hline Wireless channel data rate & 54Mbps \\
\hline Receive threshold & -80dBm \\
\hline UAV energy consumption & 174.21W\tablefootnote{This parameter is based on the specifications of DJI Phantom 4.} \\
\hline MAC protocol & 802.11 \\
\hline Antenna & Omnidirectional antenna \\
\hline Simulation time & 200s \\
\hline
\end{tabular}
\end{center}
\end{table}

\subsection{Evaluation of Small-Group Mobile Multicasting}
In the small-group simulation, there are 9 UAVs among which four UAVs are receivers. Among the four receivers, one UAV is a mobile receiver. This mobile UAV receiver moves a distance of 102.6 meters at the speed of 10m/s during the multicasting. We vary the network traffic load from 512Kbit/s to 2.176Mbit/s to observe the AMD, AMT, and AMoT performance for the five schemes.
\begin{figure}[h]
\begin{center}
\begin{tabular}{c}
\includegraphics[trim=20 180 40 180,clip,height=2.4in]{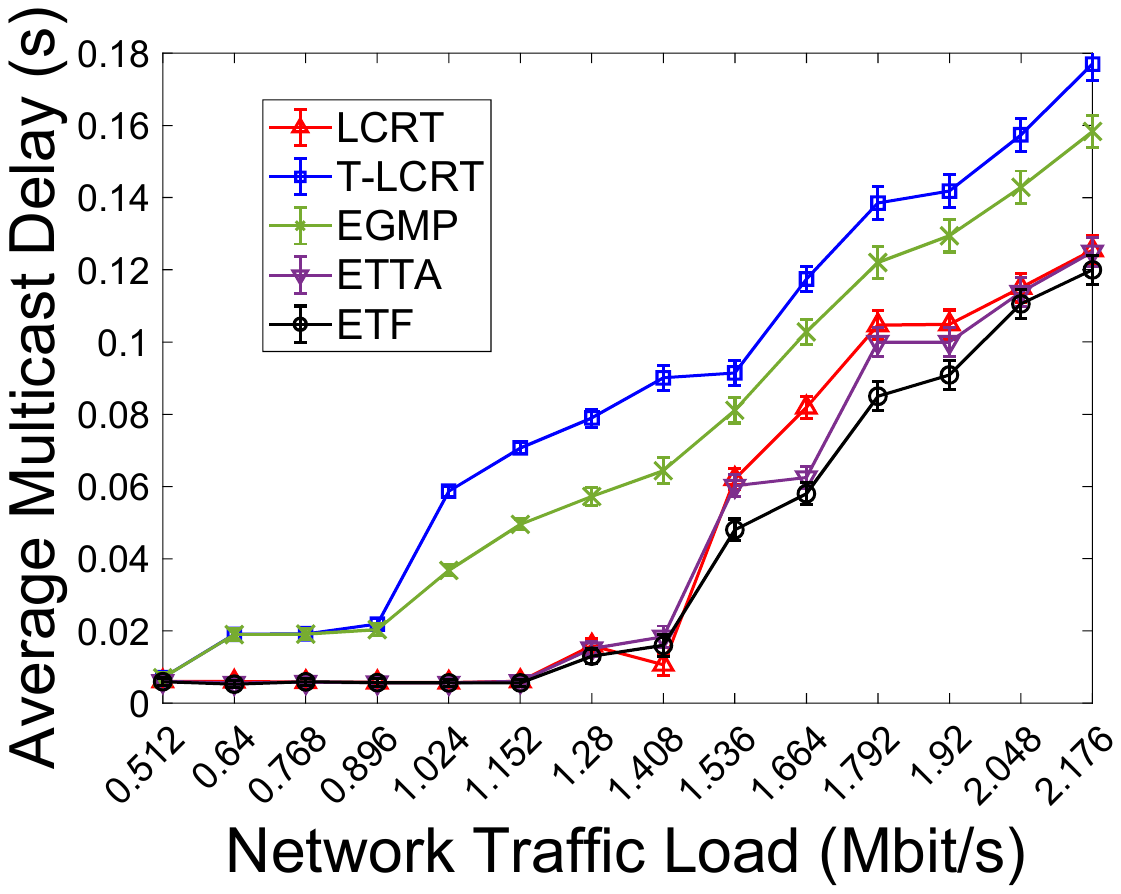}
\end{tabular}
\end{center}
\caption{Comparison of the average multicast delays in the small-group simulation.}\label{smallgroupdelay}
\end{figure}
\begin{figure}[h]
\begin{center}
\begin{tabular}{c}
\includegraphics[trim=20 180 40 200, clip,height=2.4in]{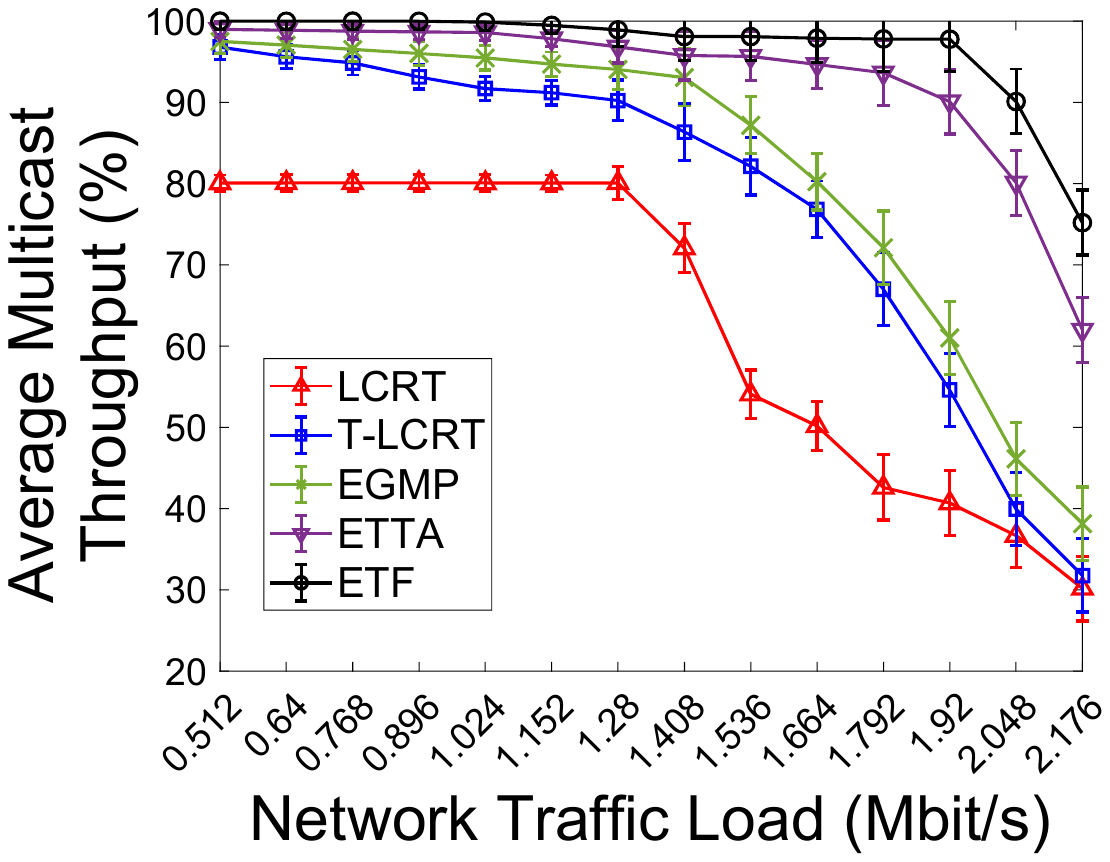}
\end{tabular}
\end{center}
\caption{Comparison of the average multicast throughput in the small-group simulation.}\label{smallgroupthroughput}
\end{figure}
\begin{figure}[h]
\begin{center}
\begin{tabular}{c}
\includegraphics[trim=20 160 20 180, clip,height=2.4in]{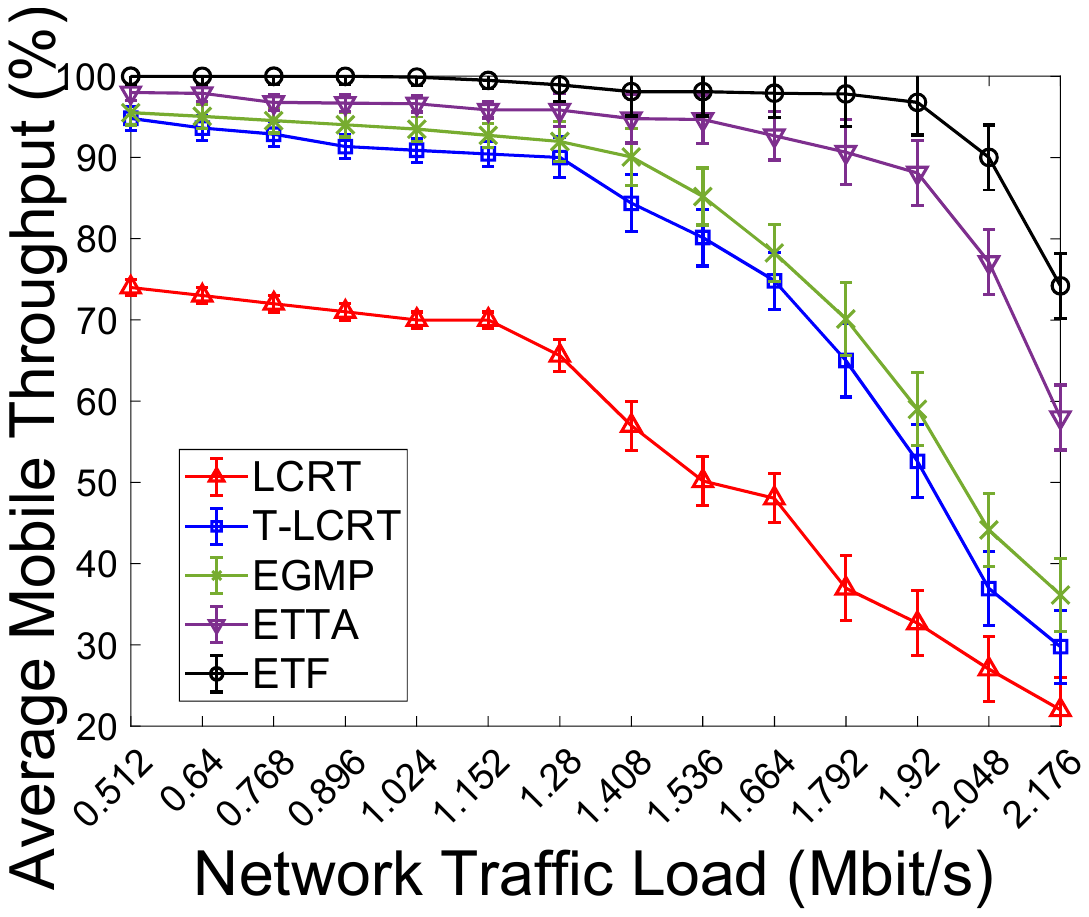}
\end{tabular}
\end{center}
\caption{Comparison of the average mobile throughput in the small-group simulation.}\label{smallmobilethroughput}
\end{figure}
\vspace{0.05in}
\begin{figure}[h]
\begin{center}
\begin{tabular}{c}
\includegraphics[trim=20 20 20 20, clip,height=2.2in]{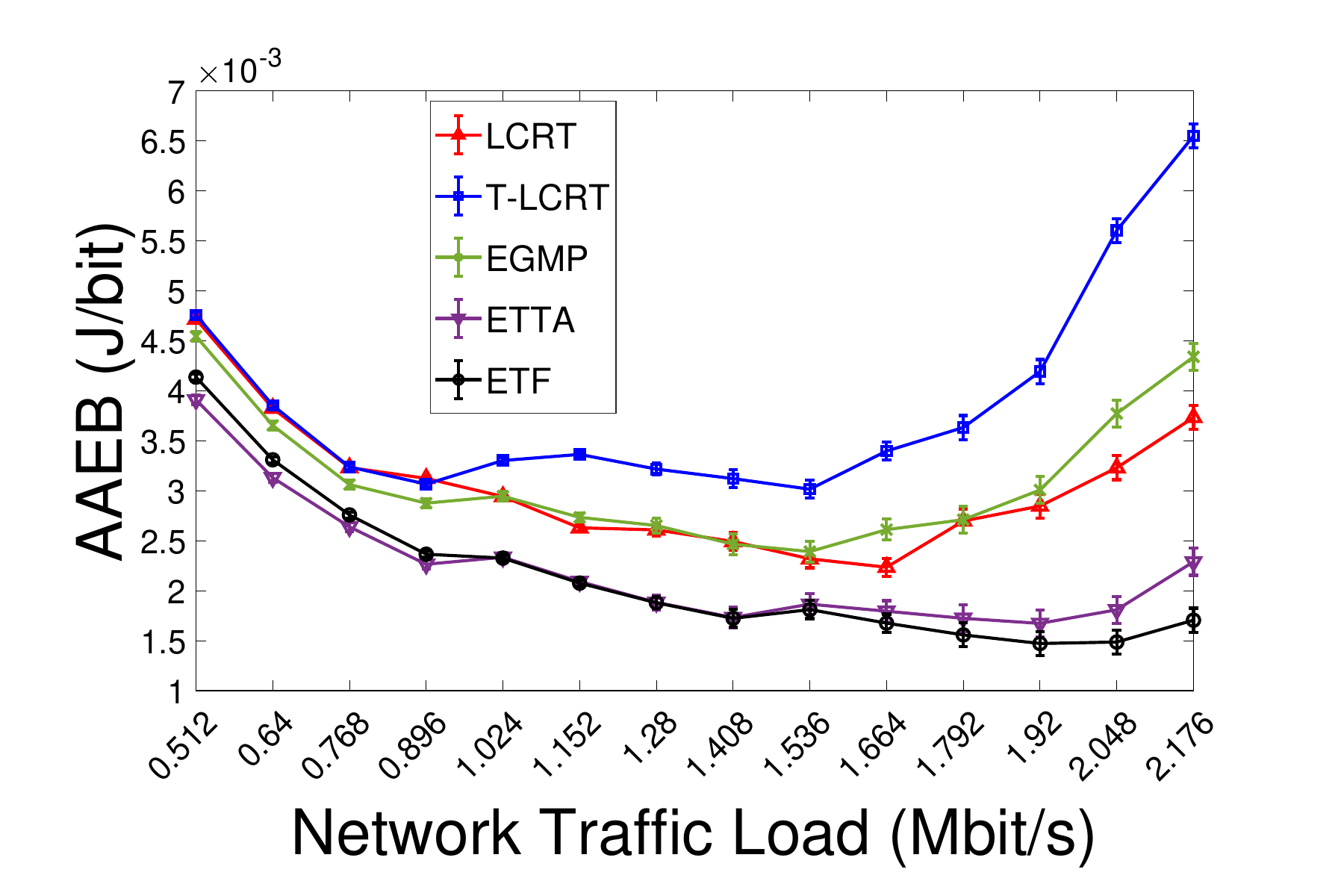}
\end{tabular}
\end{center}
\caption{Comparison of the average additional energy consumption per received bit in the small-group simulation.}\label{smallmobileenergy}
\end{figure}
\vspace{0.05in}

Fig.~\ref{smallgroupdelay} shows the AMD performance. LCRT, ETTA, and ETF achieve shorter AMDs than T-LCRT and EGMP do. This is because T-LCRT and EGMP employ nodes that are not forwarders on the multicasting structure to be transition forwarders for the mobile UAV, while ETF and ETTA make use of forwarders on the LCRT tree to handover the mobile UAV and LCRT does not implement any handover process. The employment of transition forwarders that are not on the multicasting structure generates extra traffic for the system, prolonging the multicasting delays of T-LCRT and EGMP. Between T-LCRT and EGMP, T-LCRT issues control traffic to the system in order to determine suitable transition forwarders. This extra traffic load worsens AMD for T-LCRT as compared to EGMP. LCRT, ETTA and ETF achieve all AMDs under 150ms in this simulation. The AMD difference is due to the fact that they calculate AMDs based on different packets: LCRT drops the greatest number of packets and its AMDs do not take those dropped packets into account, ETTA drops some packets (because the established trajectory may not always be seamless) and its AMDs are based on received packets, and ETF calculates the AMDs based on all transmitted packets.

\vspace{0.05in}

Fig.~\ref{smallgroupthroughput} plots the AMT performance. ETF achieves the highest AMTs of the five simulated protocols. This is because ETF attempts to transit the mobile UAV seamlessly by planning trajectories that are always covered by the LCRT tree. ETF trajectories are formed by forwarders on the LCRT tree and hence no extra data traffic is generated for the system. Also, such trajectories are planned based on knowledge of LCRT forwarders' coordinates which have obtained while establishing the LCRT tree. Therefore, ETF also generates little control traffic to the system. ETTA adopts the similar strategy.  However, its trajectory may not always be seamless which causes some data dropped. As traffic load increases, so does the AMT difference between ETTA and ETF, mainly because of the combined effect of data loss in ETTA and the gradually increased interference in the multicasting system. LCRT has the lowest AMT because the mobile UAV does not receive any data during its movement. EGMP and T-LCRT both employ transition forwarders to provide connections to the mobile UAV when this UAV is away from its original forwarder's RTR, allowing them to achieve better AMT than LCRT. Moreover, EGMP determines transition forwarders without changing the multicast structure. EGMP transition forwarders can hence transmit to the mobile UAV in time, a major contributing factor to EGMP's higher AMT than T-LCRT.

\vspace{0.05in}

The average throughput results achieved by the mobile UAV with the five multicasting protocols are plotted in Fig.~\ref{smallmobilethroughput}. In our simulation, ETF can guarantee the receipt of acceptable throughput performance for the mobile UAV until the network traffic load is larger than 1.92Mb/s. The reason for the great throughput degradation when the traffic load becomes greater than 2.048Mb/s is mainly because there is only one channel in the multicasting system. When transmission rates increase, the interference between multicasting paths or between different hops on the same path become more intensive, degrading the throughput performance. The throughput of the ETTA-based mobile UAV is a bit worse than ETF's performance because the ETTA-based UAV may lose some data during its transition. EGMP and T-LCRT can guarantee acceptable AMoTs when the network traffic loads are not greater than 1.408Mb/s and 1.28Mb/s respectively. With LCRT, the mobile UAV does not receive data after being disconnected from the original forwarder and before moving into another LCRT forwarder's RTR.

\vspace{0.05in}

Fig.~\ref{smallmobileenergy} plots the AAEB performance from small-group simulations. T-LCRT consumes more additional energy per received bit than other protocols because it has the longest transition paths and its throughput (i.e., the number of received bits) is lower than EGMP, ETTA, and ETF. The AAEBs of T-LCRT decrease when the transmission rate initially increases mainly because its number of received bits increases but the moving trajectory remains the same (and so the consumed energy) at this time period. However, when transmission rates continuously grow, with the same transmission power, the RTR shrinks (incurring longer transition paths and more control traffic) and the throughput decreases, giving the major reason for the increase in AAEB of T-LCRT. A similar explanation can justify the AAEB curve of EGMP. As for LCRT, the mobile UAV transits via the same path at different transmission rates. Its AAEBs are mostly affected by the number of received bits at different transmission rates. Between ETF and ETTA, when the transmission rates are not greater than 1.024Mbit/s, they receive a similar number of bits but ETTA has a slightly shorter transition path because of the different ways to determine $T$; when the transmission rates are greater than 1.024Mbit/s, the number of received bits by ETTA is obviously smaller than the number of bits received by ETF, yielding a higher AAEB performance.

\vspace{0.05in}

As shown in the second column of Table III, we also observe the extra control traffic generated by EMGP, T-LCRT, ETTA, and ETF in this small-group simulation. ETTA and ETF generate less additional control traffic than EGMP and T-LCRT. The additional control traffic of ETTA and ETF is mainly composed of multicasting forwarders' coordinates and referred coverage radii that are piggybacked on packets exchanged between multicasting forwarders while establishing the multicasting tree. EGMP's additional control traffic not only includes the information of zone leaders' coordinates and zone IDs but also packets issued to seek a zone leader that may support this mobile transition. T-LCRT generates the most additional control traffic as it needs to search for transitional forwarders by enabling communications between potential transitional forwarders. LCRT is not included in the table because it does not support mobile UAV transitions and hence does not generate additional control traffic for this purpose. Overall, in the small-group simulation, ETF performs better than other schemes while consuming less energy and issuing less extra traffic to the multicasting system.
\begin{table}[h]
\begin{center}
\caption{Control traffic generated in the small-group or large-group simulations.} \label{parameters1}\vspace{1em}
\begin{tabular}{|l|l|l|}
\hline {\bf Multicasting schemes} & \multicolumn{2}{c|}{\bf Control traffic (Kbits)} \\ \cline{2-3}
& In the small-group & In the large-group \\
& simulations & simulations \\ \cline{2-3}
\hline EGMP & 2.152 & 9.36 \\
\hline T-LCRT & 6.16 & 20.112 \\
\hline ETTA & 0.864 & 1.632 \\
\hline ETF & 0.864 & 1.632 \\
\hline
\end{tabular}
\end{center}
\end{table}

\subsection{Evaluation of Large-Group Mobile Multicast}
\begin{figure}[h]
\begin{center}
\begin{tabular}{c}
\includegraphics[trim=20 180 40 200, clip,height=2.4in]{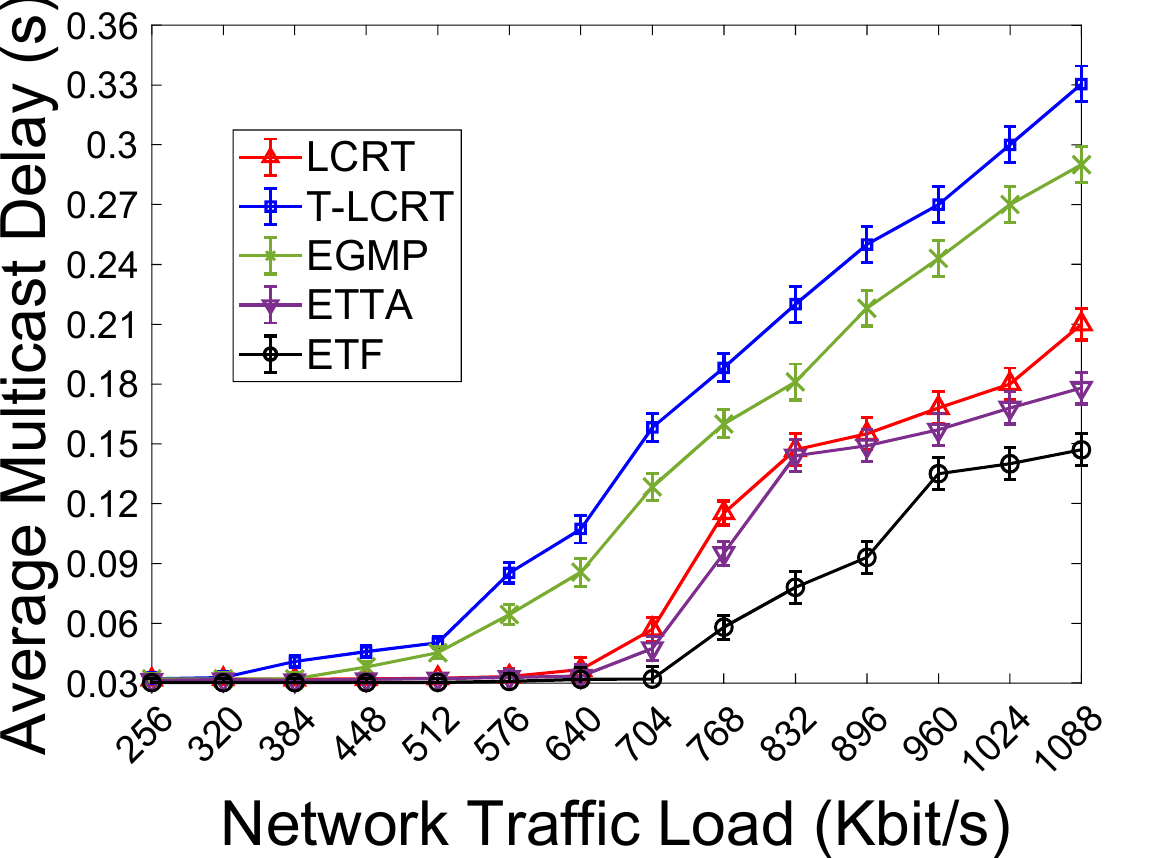}
\end{tabular}
\end{center}
\caption{Comparison of the average multicast delays in the large-group simulation.}\label{largegroupdelay}
\end{figure}
\begin{figure}[h]
\begin{center}
\begin{tabular}{c}
\includegraphics[trim=20 180 40 200, clip,height=2.4in]{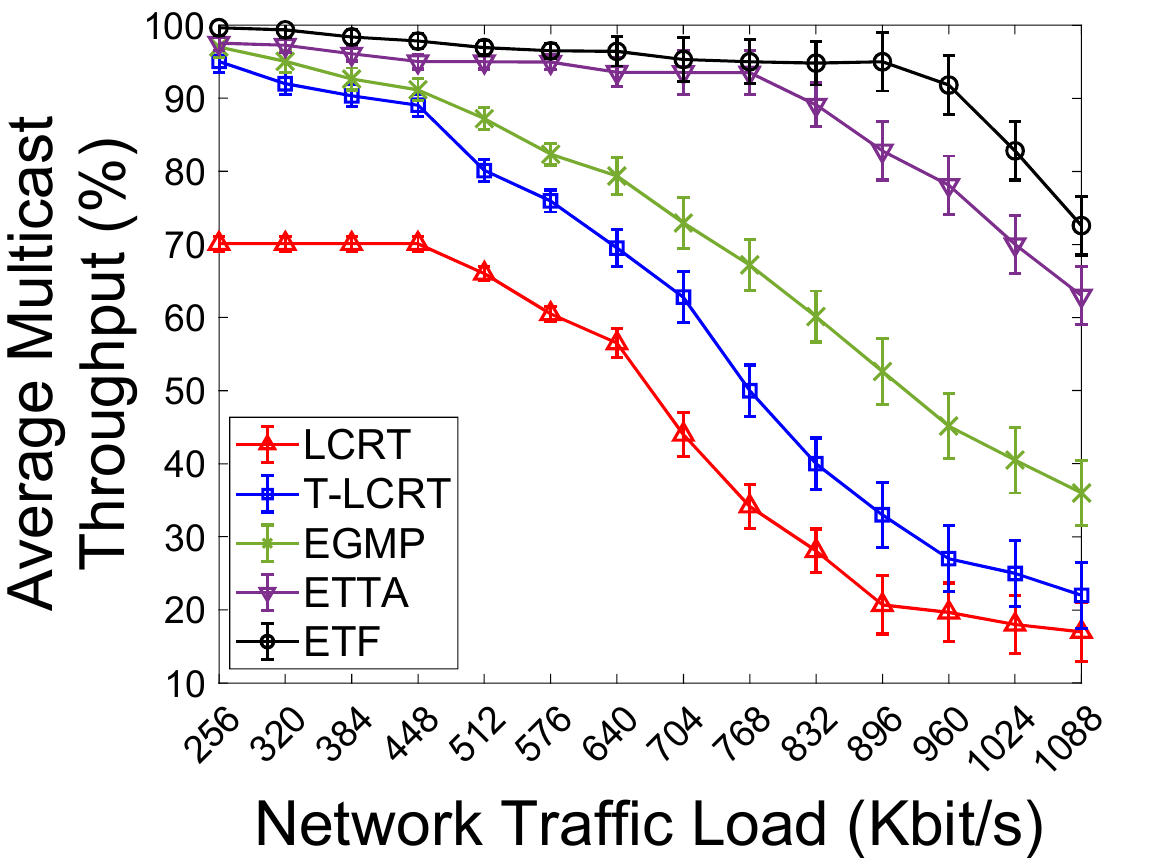}
\end{tabular}
\end{center}
\caption{Comparison of the average multicast throughput in the large-group simulation.}\label{largegroupthroughput}
\end{figure}
\begin{figure}[h]
\begin{center}
\begin{tabular}{c}
\includegraphics[trim=20 180 40 200, clip,height=2.4in]{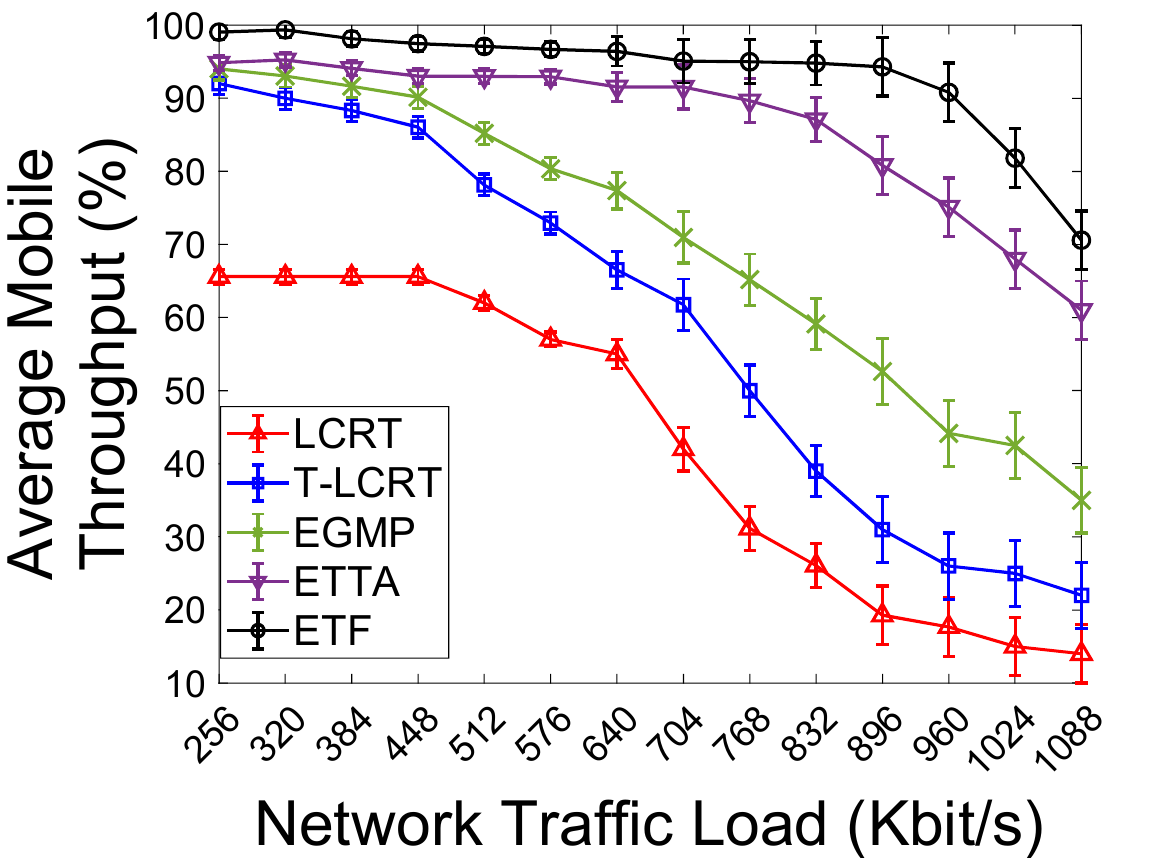}
\end{tabular}
\end{center}
\caption{Comparison of the average mobile throughput in the large-group simulation.}\label{mbthroughputlarge}
\end{figure}
\begin{figure}[h]
\begin{center}
\begin{tabular}{c}
\includegraphics[trim=20 20 20 20, clip,height=2.3in]{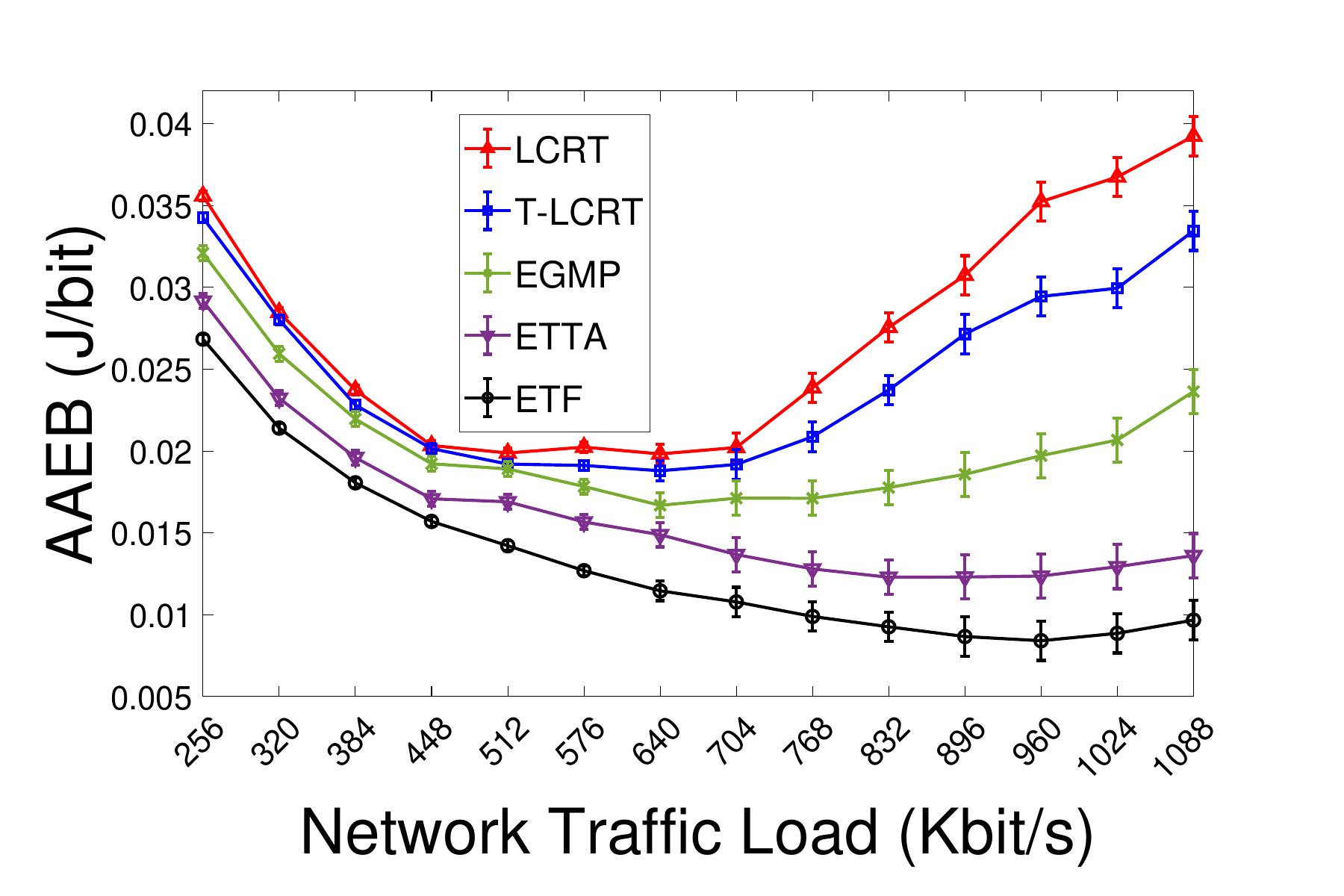}
\end{tabular}
\end{center}
\caption{Comparison of the average additional energy consumption per received bit in the large-group simulation.}\label{largegroupenergy}
\end{figure}

In the large-group simulation, the 165 UAVs are distributed so that each RTR has 10 UAVs. There are 3 mobile UAVs: the first one moves 587 meters at a speed of 10m/s, the second one moves 346 meters at a speed of 25m/s, and the third one moves 608 meters at a speed of 20m/s. We evaluate the four multicasting algorithms when the network traffic load varies from 128Kbit/s to 960Kbit/s. Based on Fig.~\ref{largegroupdelay}, ETTA, LCRT, T-LCRT, and EGMP yield similar relative results for AMDs in the large-group simulation as was observed for the small-group simulation. Similar reasons for the results in Fig.~\ref{smallgroupdelay} can explain the results of the four protocols plotted in Fig.~\ref{largegroupdelay}. In contrast, ETF achieves clearly shorter delays than  the other four protocols in the large-group simulation. This is because ETF always establishes seamless trajectories for mobile UAVs. But ETTA and LCRT may cause lost packets. When transitions happen closer to the multicast source, the packets sent to these transition UAVs should have shorter delays which may however be lost in ETTA an LCRT, generating longer AMDs.

\vspace{0.1in}

Fig.~\ref{largegroupthroughput} shows that ETF achieves the highest AMTs in the large-group simulation. It achieves good AMT ($90\%$ or above) when the network traffic load is not larger than 980Mb/s, while ETTA achieves good AMTs when the network traffic load is less than 780Mb/s. This shows that, in this simulation, to achieve acceptable AMTs, ETF can admit 25.6\% more traffic than ETTA. This is mainly because ETF always seeks fully seamless trajectories while a few ETTA trajectories may not be seamless by our previous method. As for EGMP and T-LCRT, in the large-group simulations, they are more complicated procedures and so take more time to find transition forwarders. More transition forwarders issue extra data traffic for the system as well. These factors together affect their AMTs. Moreover, in this group of simulations, there are 3 mobile UAVs transiting heterogeneously. Fig.~\ref{mbthroughputlarge} compares the average throughput of the three mobile UAVs. Similar to the reasons for the results in Fig.~\ref{largegroupthroughput}, all mobile UAVs achieve better throughput performance with ETF than when transitting with ETTA, EGMP, T-LCRT, and LCRT.

Fig.~\ref{largegroupenergy} shows the average additional energy consumption per received bit (AAEB) for the five protocols in the large-group simulation. LCRT consumes more energy than other compared protocols because the number of received bits with LCRT is the lowest. Also, in the large-group transitions, the ratios of the distances of LCRT trajectories to the distances of T-LCRT trajectories is smaller as compared to those in the small-group transitions. As for T-LCRT and EGMP, T-LCRT transitions consume more energy than those for EGMP. T-LCRT forms its transition paths by looking into connections between nodes while EGMP adjusts its transition paths between zones, enabling EGMP to have shorter transition paths and less control message exchanges. ETTA outperforms T-LCRT and EGMP because it attempts to employ the shortest transition paths, does not have additional control message exchanges, and has a higher throughput. Regarding ETF, it is able to use a long-distance SLT to transit UAVs if the SLT is seamless. ETF also receives the greatest number of bits. These enable ETF to have the lowest AAEBs in this simulation.

The extra control traffic generated by each multicasting scheme in this simulation is in the third column of Table III. As compared to the small-group simulation, the increase of control traffic is the least in ETTA because such increase is mainly caused by the fact that more multicasting forwarders in the large-group simulation exchange their location information while establishing its multicasting architecture than those in the small-group simulation. Similarly, the large-group EGMP also has more zone leaders to exchange location and ID information. Moreover, as there are more mobile UAVs to transit and they move longer distances than the mobile UAV in the small-group simulation, more control communications between potential transitional forwarders have been carried out in the large-group EMGP simulation. These cause EGMP to generate more control traffic than ETTA in the large-group simulation. For T-LCRT, its ACO increases the most among the three multicasting schemes. This is mainly because there are many more UAVs involved in supporting mobile UAV transitions. These UAV are not only multicasting forwarders but also multicasting receivers or UAVs that are not already on the multicasting tree. These UAVs need to communicate to decide transitional forwarders. Overall, combining both AMDs and AMTs, ETF can admit up to 66\% more traffic than EGMP, T-LSRC, and LCRT. This improvement is achieved with the smallest energy consumption per received bit as well as the lowest control traffic generated.

\subsection{Evaluation of Mobility Impact}
Like the large-group simulation, there are 165 UAVs in this simulation. However, we change the locations of some UAVs, enabling the multicasting schemes to form multicasting architectures different from those established in the large-group simulation. As such, multicasting paths with more variable numbers of hops between each other are formed, allowing mobile UAVs to have more different transition activities. We vary the number of mobile UAVs from 1 to 12 to observe the performance of AMoDs and AMoTs for the five multicasting schemes. These mobile UAVs start transiting at different times, move at different speeds which are in the range $[10, 35]$m/s, and have their origins and destinations located in two overlapping forwarders, two short-distance non-overlapping forwarders, and two long-distance non-overlapping forwarders. The multicasting data rate is 256Kbit/s.
\begin{figure}[h]
\begin{center}
\begin{tabular}{c}
\includegraphics[trim=20 180 40 200, clip,height=2.4in]{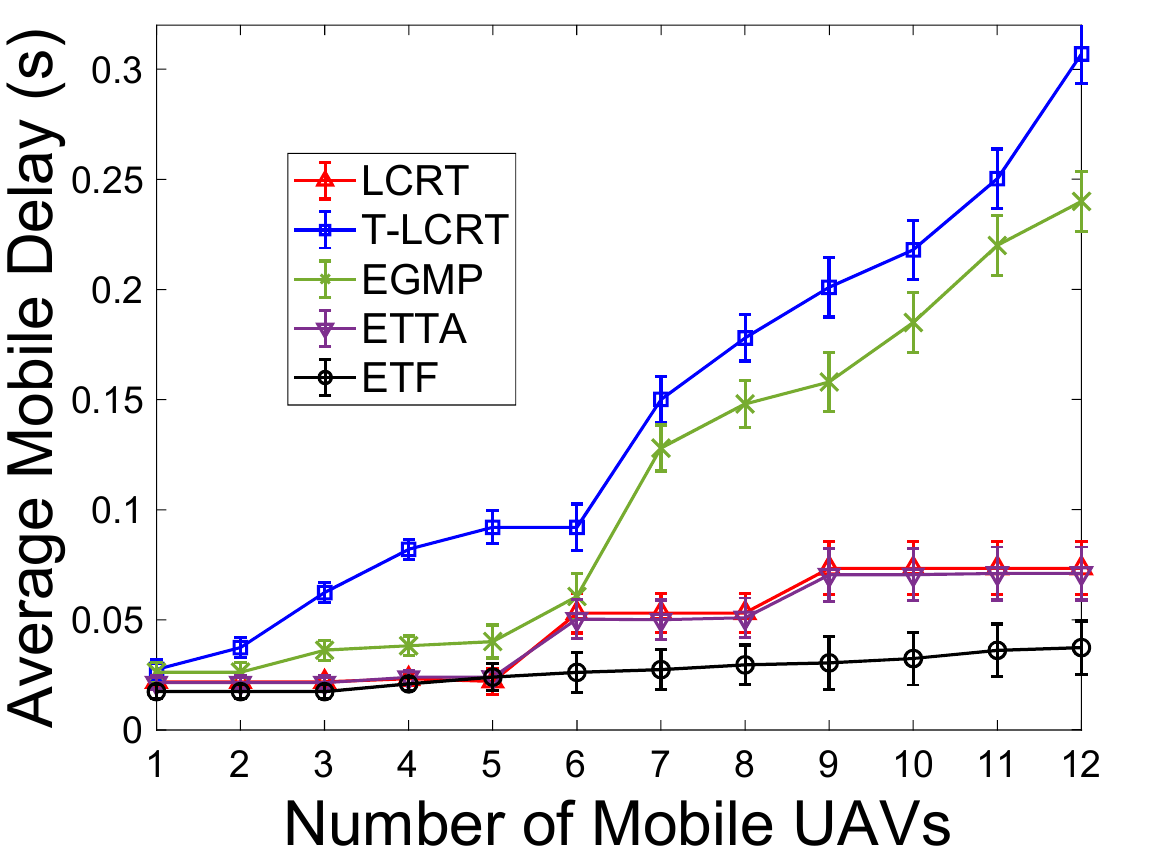}
\end{tabular}
\end{center}
\caption{Comparison of the average mobile delays when the number of mobile UAVs varies from 1 to 12.}\label{mobilitydelay}
\end{figure}

Fig.~\ref{mobilitydelay} plots the AMoDs achieved by the five multicasting schemes. T-LCRT and EGMP generate longer AMoDs than ETF, ETTA and LCRT do. This is because T-LCRT and EGMP issue control traffic to the multicasting system in order to seek transition forwarders for mobile UAVs. Also, the transition forwarders employed by T-LCRT and EGMP may not be forwarders on their multicasting architectures, causing more UAVs to compete for wireless channels in order to forward data. Between T-LCRT and EGMP, T-LCRT seeks transition forwarders among all UAVs in the system while EGMP selects transition forwarders from zone leaders. T-LCRT hence issues more control traffic to the system when determining suitable transition forwarders, prolonging the delays for mobile UAVs to receive data as compared to EGMP. ETF achieves much shorter AMoDs than T-LCRT and EGMP because ETF can determine transition forwarders without issuing much extra control traffic and also it only employs those UAVs that are currently forwarding multicasting data as transition forwarders. The slight increase in ETF's AMoDs when the number of mobile UAVs increases is mainly because those mobile UAVs transiting later connect to the multicasting sender via a path with more hops. For ETTA and LCRT, its AMoDs are obtained based on packets excluding those transmitted during the transition interruptions. Plus no additional traffic generated for transiting mobile UAVs, the AMoDs of LCRT and ETTA are shorter than other multicasting schemes. The increments of AMoDs of ETTA and LCRT when there are 5 and 8 mobile UAVs in the system are because the 5th and the 8th mobile UAVs are close to the multicasting source and the dropped packets due to their transitions should have short delays.
\begin{figure}[h]
\begin{center}
\begin{tabular}{c}
\includegraphics[trim=20 180 40 200, clip,height=2.4in]{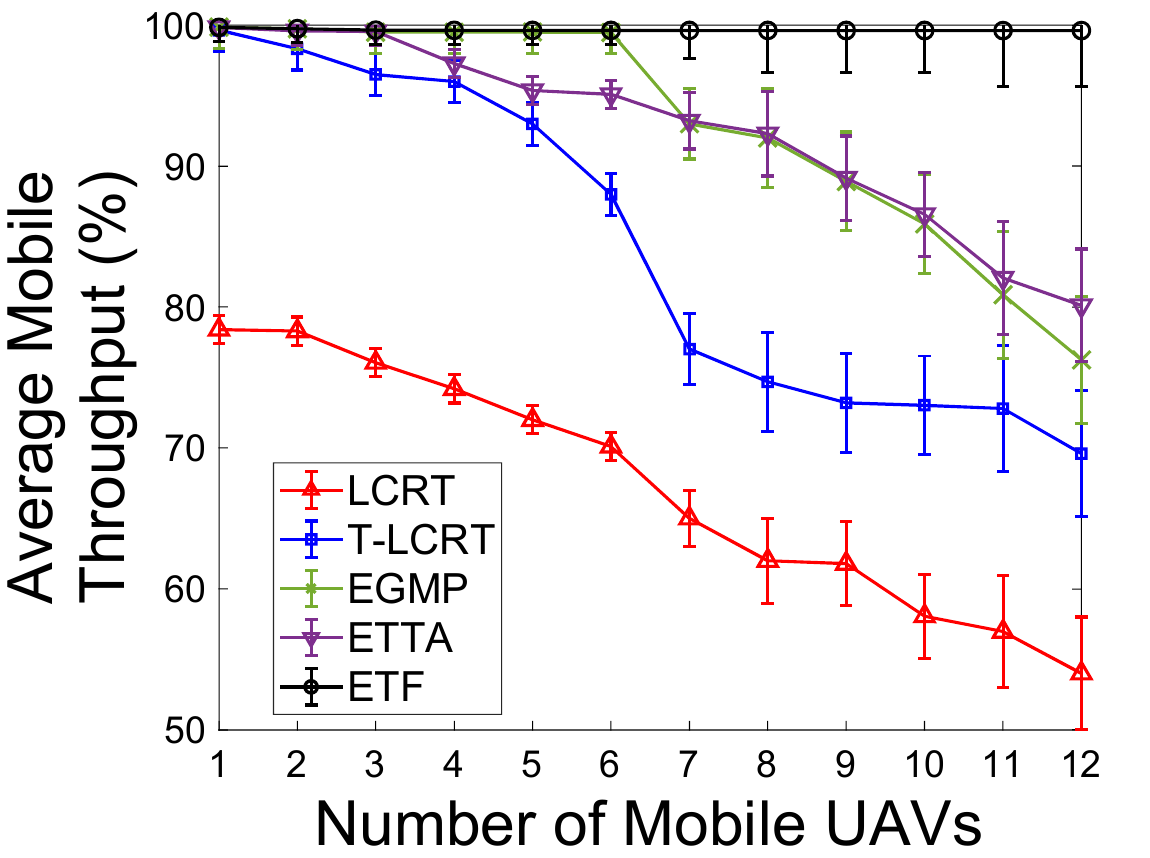}
\end{tabular}
\end{center}
\caption{Comparison of the average mobile throughput when the number of mobile UAVs varies from 1 to 12.}\label{mobilitythroughput}
\end{figure}
\begin{figure}[h]
\begin{center}
\begin{tabular}{c}
\includegraphics[trim=20 20 20 20, clip,height=2.3in]{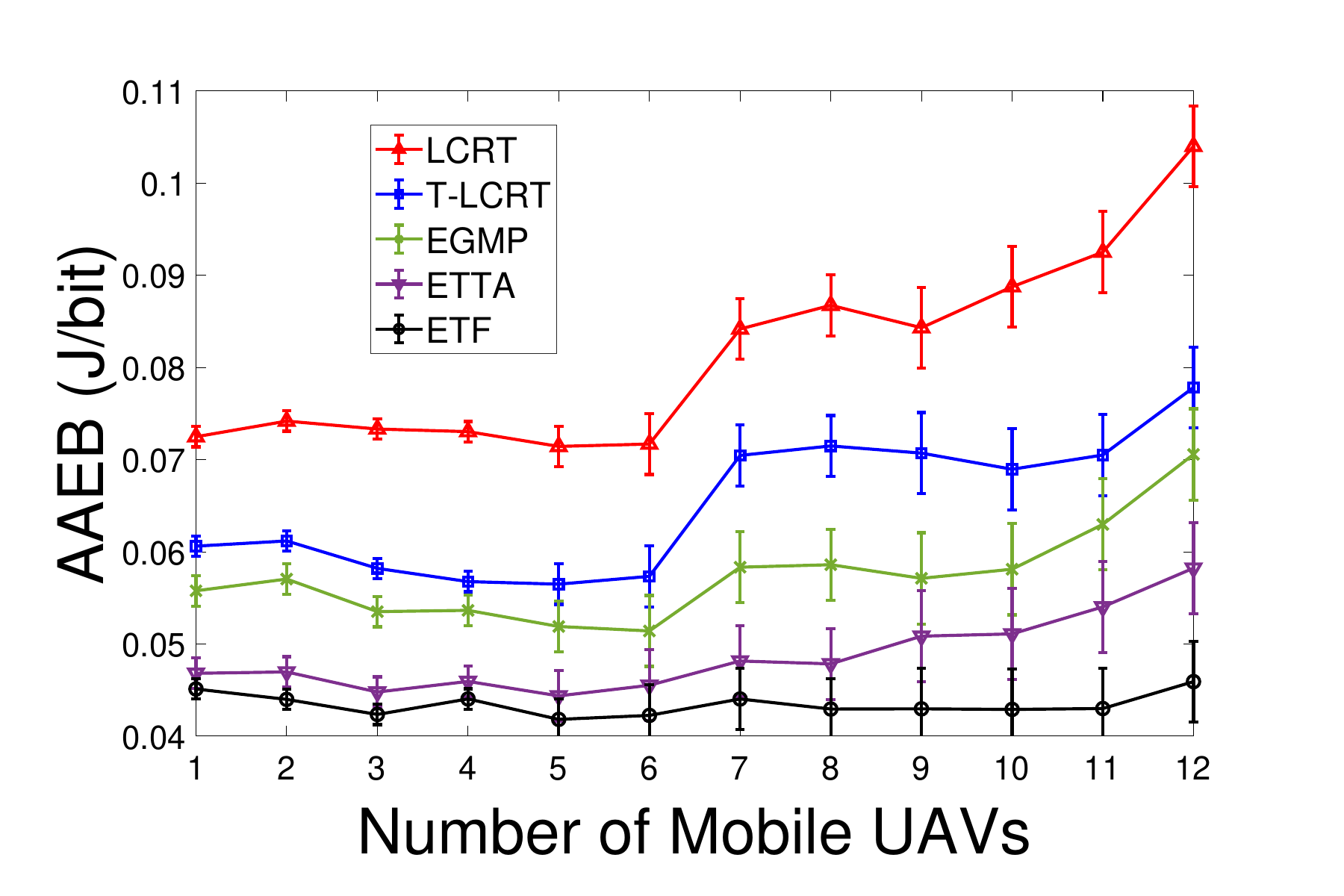}
\end{tabular}
\end{center}
\caption{Comparison of the average additional energy consumption per received bit when the number of mobile UAVs varies from 1 to 12.}\label{mobileenergy}
\end{figure}

\vspace{0.1in}

Fig.~\ref{mobilitythroughput} plots the AMoT curves. Mobile UAVs in the ETF multicasting achieve acceptable throughput performance throughout the simulation. This is mainly because ETF guarantees the seamless transitions of all mobile UAVs with limited extra control traffic load issued to the system. EGMP and T-LCRT need to seek transition forwarders and so there are some interruptions of mobile UAVs' connections to the multicasting architectures, causing lower AMoTs than that of ETF. Moreover, EGMP selects transition forwarders without changing the multicasting architecture between zone leaders. Inactive zone leaders start forwarding data when they are asked to support mobile UAV transitions. As for T-LCRT, it updates its multicasting architecture when a multicasting receiver or a non-multicasting UAV is selected as a transition forwarder. Hence, the AMoTs of EGMP are higher than those of T-LCRT. Mobile UAVs with ETTA establishes new trajectories without issuing extra traffic to the system. However, it forms a few trajectories that are not seamless,  causing data loss and hence the degraded AMoTs. For LCRT, mobile UAVs cannot receive data during the transition, incurring the lowest AMoTs among the five compared multicasting schemes. Fig.~\ref{mobileenergy} plots the AAEB performance when the number of mobile UAVs varies. As the number of mobile UAVs increases, the amount of lost data using LCRT increases greatly, generating the greatest AAEBs. The comparison between T-LCRT and EGMP is mostly explicable in the same way as in their comparisons for the first two simulations. Between ETF and ETTA, the higher throughput and the employment of SLTs for some mobile UAVs allow ETF to be more sustainable in terms of AAEBs than ETTA in our simulation.
\begin{table}[h]
\begin{center}
\caption{Control traffic generated when the number of mobile UAVs increases.} \label{parameters1}\vspace{1em}
\begin{tabular}{|l|l|l|l|}
\hline {\bf Multicasting schemes} & \multicolumn{3}{c|}{\bf Control traffic (Kbits)} \\ \cline{2-4}
& 1 mobile & 6 mobile & 12 mobile \\
& UAV & UAVs & UAVs \\ \cline{2-4}
\hline EGMP & 2.92 & 10.648 & 22.240 \\
\hline T-LCRT & 6.288 & 37.504 & 76.096\\
\hline ETTA & 1.632 & 1.632 & 1.632\\
\hline ETF & 1.632 & 1.632 & 1.632\\
\hline
\end{tabular}
\end{center}
\end{table}

Table IV gives the additional control overheads generated by EGMP, T-LCRT, ETTA, and ETF when there are 1, 6, 12 mobile UAVs in this simulation. The ACOs of EGMP and T-LCRT grow quickly as the number of mobile UAVs increases in the system, while ETTA and ETF manage to issue the same amount of ACOs for transiting a different number of mobile UAVs. This is because the ACOs of ETTA and ETF are mainly composed of the location information exchanged while establishing its multicasting architecture. For EGMP and T-LCRT, they need to seek suitable transition forwarders which generate considerable control traffic, especially when the transitions happen between two long-distance locations. We do not give the ACOs of LCRT in Table IV because LCRT does not transit mobile UAVs and hence does not generate ACOs. Overall, ETF demonstrates its scalability in achieving acceptable delay and throughput performance for mobile UAVs with less consumed energy and controlled additional control traffic.
 
\section{Conclusion}
In this paper, we studied mobile UAV multicasting in order to enable high-performance group communications between UAVs even when UAVs move their locations. Our development focused on how to seamlessly and quickly transit mobile UAVs in a resource-efficient manner, while maintaining performance-guaranteed multicasting. A new algorithm, ETF, was proposed that takes advantage of the aerial communication environment with greatly reduced obstacles to establish SLTs for transiting UAVs. As SLTs may not always be seamless, we analysed the seamlessness condition for transitions between overlapping forwarders and proposed the efficient seamlessness checking algorithm for transitions between non-overlapping forwarders. To replace an interrupted SLT, we designed the efficient trajectory formation scheme for transitions between overlapping forwarders and the efficient seamless trajectory formation algorithm for transitions between non-overlapping forwarders. In general, an ETF trajectory consists of a minimum number of seamless straight lines that intersect at the locations (derived by the efficient trajectory formation scheme) to greatly control mobile UAVs' seamless travel distances. Because the ETF algorithm employs multicasting forwarders for the seamlessness check and transition support, it allows quick transitions while controlling interference and traffic overheads. Our simulation results proved that ETF delivers multicast data with acceptable performance even when the multicasting system carries up to 66\% more traffic than the compared mobile multicasting protocols. Since ETF does not require a wireless multicasting algorithm to change its procedure, it can be easily integrated with an existing wireless multicasting algorithm to support high-performance yet resource-efficient UAV transitions.

\bibliographystyle{plain}

\end{document}